\documentclass[12pt,preprint]{aastex}

\shorttitle{$\alpha$-ENHANCED LICK SPECTRAL INDICES}
\shortauthors{H.-c. LEE \& G. Worthey}

\begin{document}

\title{$\alpha$-ENHANCED INTEGRATED LICK/IDS SPECTRAL INDICES\\
       AND MILKY WAY AND M31 GLOBULAR CLUSTERS\\
       AND EARLY-TYPE GALAXIES}

\author{Hyun-chul Lee \& Guy Worthey}
\affil{Department of Physics and Astronomy, Washington State University, 
    Pullman, WA 99164-2814}

\begin{abstract}
All 25 Lick/IDS spectral indices have been computed for the integrated
light of simple stellar populations over broad ranges of age and
metallicity and with effects from horizontal-branch stars fully
implemented. Our models employ $\alpha$-enhanced isochrones at the
sub-solar metallicity regime, but solar-scaled ones at solar and
super-solar metallicity.  We have also employed the updated response
functions of Houdashelt et al. at the solar and super-solar
metallicity regime, so that we could assess the light-element
enhancement phenomena seen from metal-rich early-type galaxies. For
Balmer indices a significant response was noted for H$\gamma$ and
H$\delta$ when $\alpha$-elements are enhanced, but H$\beta$ is rather
$\alpha$-$\it{in}$sensitive.  We also find that our 5 and 12 Gyr
models of H$\gamma$ and H$\delta$ overlap in the metal-poor regime
because of changing populations of blue horizontal-branch
stars. Furthermore, for populations younger than 1 Gyr, Balmer lines
become weaker in the metal-poor regime because the main-sequence
turnoff is hotter than 10,000 K.  We present models at fixed [Fe/H]
(rather than fixed heavy element mass fraction $Z$) and compare to
Milky Way globular clusters that have independently estimated mean
[Fe/H] and [$\alpha$/Fe]. Comparison of our models with observations
of Milky Way and M31 globular clusters in index-index space are
favorable, tracing the observations at a model age of 12 Gyr {\em
without any zero-point shifts} that are needed by some other models.
The metallicity range of M31 globular clusters is similar to that of
their Galactic counterparts.  We also verify Beasley et al.'s recent
hypothesis of the existence of young and intermediate-age star
clusters in M31.  Contrary to the literature values, the Milky Way
globular cluster NGC 6553 appears more metal-rich than NGC 6528 from
metal indices. We present H$\delta$ and H$\gamma$ Lick/IDS indices for
the Lick/IDS sample of galaxies.  We confirm the well-known
enhancement of Mg and Na relative to Fe and Ca among 
early-type galaxies, and its increase with increasing velocity
dispersion.  There are distinct differences between globular clusters
and galaxies in diagrams involving CN$_{1}$ and CN$_{2}$, hinting that
the globular cluster environment may be a special one in terms of the
amount of N incorporated into stars. 

\end{abstract}


\section{INTRODUCTION}

Lick IDS spectral feature indices \citep{wfgb94} are measurements of
the depths of some of the more obvious absorption blends in the
spectra of stars and galaxies. They have been widely used to derive
the mean age and mean chemical compositions of stellar systems such as
star clusters and galaxies. Models for these indices, when compared
with observations, can shed light on the ages and chemical
compositions of objects as small (and local) as globular clusters to
objects as large (and cosmological) as giant elliptical galaxies. As
new model ingredients become available it is essential to continue to
validate the use of the integrated Lick IDS spectral indices for age
and metallicity estimation, especially for distant, unresolved stellar
systems.  One way is through the satisfactory confirmation of age and
metallicity of the Milky Way globular clusters (GCs) with their
independently acquired age and metallicity (e.g., \citealt{gib99}).

In order to compute the theoretical integrated Lick IDS spectral
indices for simple stellar populations (SSPs; clusterlike populations
characterized by a single age and a single abundance at birth), two 
components are generally needed. They are (1) stellar evolutionary 
tracks or isochrones and (2) some connection to observable quantities 
such as synthetic stellar fluxes or empirically derived index fitting
functions.  The fitting functions of \citet{wfgb94}\footnote{ See also
\citet{wor97} for H$\delta_{A}$, H$\gamma_{A}$, H$\delta_{F}$,
H$\gamma_{F}$.} have been built upon the Lick stellar library of Milky
Way stars.  Because of the empirical nature of the fits, we expect
that the fitting functions follow the abundance pattern of the stars
to which they were fit. That is, in the metal-poor regime ([Fe/H] $<$
0.0) they should represent stars of [$\alpha$/Fe] $\sim$ +0.3 dex and
at the solar and super-solar metallicity regime they represent stars
of [$\alpha$/Fe] $\approx$ 0.0.

Bearing this in mind, we employ $\alpha$-enhanced isochrones at
subsolar metallicities (in this work $Y^{2}$ isochrones,
\citealt{kim02}) in order to make SSP models that are more compatible
with the $\alpha$-element enhanced fitting functions.  By the same
token, we compute our models using solar-scaled $Y^{2}$ isochrones in
the solar and super-solar metallicity regime. Also, in order to make a
straightforward comparison with observational data, we present our
model results as a function of [Fe/H] instead of as a function of total
metallicity, [Z/H] (cf. \citealt{tra00}; \citealt{tmb03a}; hereafter
TMB03a).

Besides basic checking of our models with wide variety of data, some
of the issues that we are addressing in this study 
are the following: (1) the sensitivity of the bluer Balmer indices,
H$\gamma$ and H$\delta$, to $\alpha$-enhancement may be higher than
for H$\beta$ (\citealt{tmk04}), (2) young and intermediate-age star
clusters may be present in M31 (Beasley et al. 2004, 2005;
\citealt{bur04}), (3) there may be CN anomaly between GCs and
early-type galaxies (\citealt{bur84}), (4) the overall Mg enhancement
in early-type galaxies is seen to depend on velocity dispersion
(\citealt{wor98}, \citealt{wor92}), and (5) there may be a Ca underabundance in
early-type galaxies compared to the solar mixture (\citealt{tmb03b};
\citealt{cen04}).

In the following section, we describe our models in detail.  In $\S$ 3
we compare models with recent observations of Milky Way and M31
globular clusters and early-type galaxies.  Section 4 summarizes and
concludes our work.

\section{STELLAR POPULATION MODELS}

The present models are direct descendents of an evolutionary
population synthesis code that was developed to study the stellar
populations of globular clusters and early-type galaxies
(\citealt{lee00}; \citealt{lee02}).  In this work, we have taken
advantage of a recent emergence of the $Y^{2}$ isochrones with an
$\alpha$-mixture [this mixture has O, Ne, Na, Mg, Si, P, S, Cl, Ar,
Ca, and Ti enhanced and Al and Mn depressed relative to Fe, C, N, K,
Cr, and Ni: cf. \citet{kim02}]. This mixture may not track the halo
abundance exactly, but it should be fairly closely compatible with the
stellar library that was the basis of \citet{wfgb94}'s fitting
functions, at least at the metal-poor end. We assume that the the
behavior of the library is roughly: [$\alpha$/Fe] = +0.3 for [Fe/H]
$<$ $-$1.0, [$\alpha$/Fe] goes from +0.3 to 0.0 as [Fe/H] does from
$-$1.0 to 0.0, and [$\alpha$/Fe] = 0.0 for [Fe/H] $>$ 0.0 (e.g.,
\citealt{whe89}).

Following earlier work (Lee et al. 2000, 2002), we take into account
the detailed systematic variation of horizontal-branch (HB) morphology
with age and metallicity. The post-red giant branch evolutionary
tracks by \citet{yi97} are coupled to the $Y^{2}$ isochrones.  In
order to reproduce the observational HB morphology of Milky Way
globular clusters, a value of $\eta = 0.6$ was taken, where $\eta$ is
the scaling factor in the \citet{rei75} formula for stellar mass loss.
The value of the helium enrichment parameter, $\Delta${\it
Y}/$\Delta${\it Z} = 2, was assumed. The standard \citet{sal55}
initial mass function was adopted for calculating the relative number
of stars along the isochrones. The investigated age range is from 1 Gyr 
to 12 Gyr and the metallicities cover $-2.5 \leq$ [Fe/H] $\leq +0.5$.

The strength of spectral line indices is calculated with either 
\begin{equation}
EW = \Delta\lambda (1 - (F_{\lambda}/F_{C}))
\end{equation}
or
\begin{equation}
Mag = - 2.5 log (F_{\lambda}/F_{C}),
\end{equation}
where $\Delta\lambda$ is an index bandpass and $F_{\lambda}$ and
$F_{C}$ are the flux in the index bandpass and the pseudocontinuum
flux in the index bandpass (\citealt{wfgb94}; \citealt{wor97}). First,
for each star (or bin of stars along an isochrone) with a given
metallicity, temperature, and gravity, we find the average flux values
within both pseudocontinuum regions for each index from the stellar
model atmospheres. We have employed the stellar model atmospheres of
\citet{lej97,lej98} in this study to place $F_{C}$. We derive $F_{C}$ 
at the center of each index bandpasses by linear interpolation in 
wavelength. Second, we calculate either the equivalent width 
(EW; e.g., for H$\beta$) or the magnitude (Mag; e.g., for Mg$_{2}$) 
using \citet{wfgb94} and \citet{wor97} fitting functions, where the 
index value is given as a function of stellar [Fe/H], temperature, 
and gravity.  This allows one to solve for $F_{\lambda}$. Finally, 
we sum $F_{C}$ and $F_{\lambda}$ over all bins in the isochrone and 
compute the integrated strengths of spectral feature indices using 
equations (1) and (2).

We want to stress that our models are presented at fixed [Fe/H] so
that it can be compared directly to observed integrated cluster
spectral feature indices in the cases where the cluster has
independently-measured mean [Fe/H] and [$\alpha$/Fe].  \citet{tm03}
and \citet{mar03} have recently investigated the impact of
$\alpha$-enhanced stellar evolutionary tracks on stellar population
models using the evolution of \citet{sal00}. One should realize,
however, that because their models are provided at fixed total
metallicity [Z/H], they find that $\alpha$-enhanced tracks are hotter
than solar-scaled ones, mostly owing to the lower opacities of
$\alpha$-enhanced stellar atmospheres at fixed total metallicity
[Z/H]. The situation is opposite for models given at fixed [Fe/H]: the
$\alpha$-enhanced tracks are cooler at a given [Fe/H] (cf.  Figures 2
and 3 of \citealt{kim02}).

In order to investigate $\alpha$-enhancement at solar and super-solar
metallicities where many massive elliptical galaxies reside, we employ
scaled-solar isochrones, but include modifications to the Lick IDS
index strengths based on the updated response functions (RFs) by
\citet{hou02} which we describe in detail in the following.
                                
Recently, Houdashelt et al. (2002; hereafter HTWB02) repeated and
expanded the earlier work of Tripicco \& Bell (1995; hereafter TB95)
on the sensitivity of each Lick spectral index as 
the abundances of individual chemical elements are varied.  They used
revised and updated spectral line lists, including TiO bands that were
not included in TB95.  In addition, they added the age-sensitive
H$\gamma$ and H$\delta$ indices\footnote{The tables of spectral index
response to abundance changes are available at the
http://astro.wsu.edu/hclee/HTWB02. Also, please refer to section 3.3 
in \citet{wor04} to find the differences between TB95 and HTWB02.}.

HTWB02's experiments, like those of TB95, are done only at solar
metallicity and only for three evolutionary phases.  We apply this
treatment at the solar and super-solar regime (in our tabulated
models, the [Fe/H] = 0.0 and +0.5 entries are modified). Hence, even
though our default models at [Fe/H] = 0.0 and +0.5 have isochrones
with [$\alpha$/Fe] = 0.0, with the help from HTWB02 we can appraise
$\alpha$-element enhancement phenomena in the limit where the
abundance mixture does not feed back to change isochrone shape or
number distribution. We do not, however, apply this treatment at the
sub-solar regime because we assume that the regular fitting functions
follow the Milky Way abundance pattern.

For metal-rich indices, we include enhancement effects from the
chemical elements O, Na, Mg, Si, Ca, and Ti, while elements C, N, and
Cr track Fe in a scaled-solar fashion.  This (1) accommodates the
general observational trends seen from globular cluster stars (e.g.,
Fig. 14 of \citealt{ram03}) and (2) mimics the $\alpha$-mixture of the
$Y^{2}$ isochrones (Fig. 1 of \citealt{kim02}).  Detailed studies for
the C- and N-sensitive indices are under way and will be presented
elsewhere. Therefore, and despite the fact that we know this does not
fit all observations, the model indices CN$_{1}$, CN$_{2}$, G4300,
C$_{2}$4668 that we present in this paper assume approximately
scaled-solar behavior for C and N.

\section{COMPARISON WITH OBSERVATIONS}
    
\subsection{COMPARISON WITH MILKY WAY GLOBULAR CLUSTERS}

Having discussed the theoretical aspects of generating the integrated
$\alpha$-enhanced Lick spectral indices in $\S$ 2, we now provide
empirical checks of our models using Milky Way GCs.  In this work,
we use two datasets for the Milky Way GCs. One is from
\citet{coh98}\footnote{In this work, we use Table 2 in
\citet{bea04} who remeasured the original Cohen et al. spectra with
careful consideration of flux and wavelength calibration.  Beasley et
al.  even managed to measure some of previously unmeasured Lick
indices which reside at the shorter wavelength range, such as
H$\gamma_{A}$, H$\gamma_{F}$, Ca4227, G4300, Fe4383, Fe4531 as well as
their errors. Due to its large uncertainty we do not use NGC 6171
(M107, [Fe/H] = $-$1.04) index data. We use original C98 data
for TiO$_{2}$ since Beasley et al. did not measure this index.}
and the other is from \citet{puz02}\footnote{It is Puzia et al.'s
Table D that is used in this study. For H$\gamma$ and H$\delta$, their
Table C is used.}.  We use the \citet{har96}
compilation\footnote{http://physun.mcmaster.ca/$\sim$harris/mwgc.dat}
for [Fe/H] of the Milky Way GCs. 
Figures 1$-$6 show [Fe/H] vs. Lick indices in order to take advantage
of independent measurements of mean [Fe/H] and [$\alpha$/Fe] of the
Milky Way GCs.  

For models in Figures 1 and 2, [$\alpha$/Fe] = +0.3
dex $Y^{2}$ isochrones are employed at [Fe/H] = $-$2.5, $-$2.0,
$-$1.5, and $-$1.0, while [$\alpha$/Fe] = +0.15 dex $Y^{2}$ isochrones
are used at [Fe/H] = $-$0.5, and, finally, solar-scaled $Y^{2}$
isochrones with HTWB02 response functions of [O,Na,Mg,Si,Ca,Ti/Fe] = +0.3 dex
are applied at [Fe/H] = 0.0 and +0.5.  The effects from blue HB stars
are clear as 10 Gyr and 12 Gyr lines are compared at the metal-poor
regime. It is also noted that Balmer lines become weaker with
decreasing metallicity at 1 Gyr because the main-sequence turnoff
becomes hotter than 10,000 K in metal-poor populations.

In Figures 1-6, the data of Cohen et al. (C98, circles) are, from the
left\footnote{In Figure 2, in the upper panel for H$\gamma$, 
C98 data are,
from the left, NGC 6341, NGC 6205, NGC 6121, NGC 6838, and NGC 6356.}, and
with [Fe/H] values \citep{har96} noted in parentheses,
NGC 6341 (M92; $-$2.28), NGC 6205 (M13; $-$1.54), 
NGC 6121 (M4; $-$1.20), NGC 6838 ($-$0.73), NGC 6539 ($-$0.66), 
NGC 6760 ($-$0.52), NGC 6356 ($-$0.50), 
NGC 6624 ($-$0.44), NGC 6440 ($-$0.34), 
NGC 6553 ($-$0.21), and NGC 6528 ($-$0.04).
Those of Puzia et al. (P02, triangles) are, from the left, 
NGC 6218 (M12; $-$1.48), NGC 6626 (M28; $-$1.45), 
NGC 6981 (M72; $-$1.40), NGC 6284 ($-$1.32), NGC 6637 ($-$0.70), 
NGC 6388 ($-$0.60), NGC 6441 ($-$0.53), 
NGC 6356 ($-$0.50), NGC 6624 ($-$0.44), NGC 5927 ($-$0.37), 
NGC 6553 ($-$0.21), and NGC 6528 ($-$0.04).
There are 4 GCs in common between C98 and P02.
They are NGC 6356 ([Fe/H] = $-$0.50), NGC 6624 ([Fe/H] = $-$0.44), 
NGC 6553 ([Fe/H] = $-$0.21), and NGC 6528 ([Fe/H] = $-$0.04). 
It is useful to note that the latter two are among the most metal-rich 
Galactic globular clusters because it would be interesting to find 
if even more metal-rich GCs than these exist in other galaxies.

From Figures 1 and 2, where we compare C98 and P02 datasets of
Milky Way GCs with our Balmer line models, we find a good overall 
agreement and clear emergence of horizontal-branch morphology effects. 
We see that (1) our models of Balmer lines satisfactorily trace the 
observational data of Milky Way GCs at 12 Gyr without
zero-point shifts throughout the metallicity range, (2) C98's
most metal-poor cluster, M92 ([Fe/H] = $-$2.28) is well matched with
our models, (3) C98's M13 ([Fe/H] = $-$1.54, HB Type\footnote{HB Type
is defined as [(B $-$ R) / (B + V + R)], where B and R are the number
of HB stars left and right of the instability strip, respectively,
while V is the number of RR Lyrae stars \citep{lee94}.} = 0.97)
and P02's M12\footnote{The most metal-poor GC in the P02 sample.}
([Fe/H] = $-$1.48, HB Type = 0.92) with comparable metallicity and
similar HB morphology are rather satisfactorily reproduced from our model 
Balmer line indices, (4) C98's M4 ([Fe/H] = $-$1.20, HB Type =
$-$0.07) consistently shows strong H$\beta$, H$\gamma_{A}$, and
H$\gamma_{F}$ as expected from its HB morphology (Lee et al. 2000).

In Figures 3$-$6, 12 Gyr models for non-Balmer indices are shown. At
[Fe/H] = 0.0 and +0.5, HTWB02 response functions of [$\alpha$/Fe] =
0.0 ({\em solid lines}; our default model without HTWB02 treatment),
+0.3 ({\em dotted lines}), +0.6 dex ({\em dashed lines}) are,
respectively, employed on top of the solar-scaled $Y^{2}$ isochrones.

For each metal index we first concentrate on the sub-solar
metallicity regime to check our use of $\alpha$-enhanced $Y^{2}$
isochrones for the comparably $\alpha$-enhanced Milky Way GCs (e.g.,
\citealt{car96}).  Secondarily, we inspect the sensitivity of
integrated Lick metal line indices on $\alpha$-enhancement as we
employ HTWB02 response functions at the solar and super-solar
metallicity regime. To clarify our thinking we have divided the Lick
metal line indices into four groups. They are (1) CN-sensitive indices
CN$_{1}$, CN$_{2}$, G4300, and C$_{2}$4668, calcium-sensitive index
Ca4227, and Ca4455, (2) iron-sensitive indices Fe4383, Fe5270, Fe5335,
Fe5406, and Fe5709, (3) Mg-sensitive indices Mg$_{1}$, Mg$_{2}$, and
Mg $\it{b}$, and (4) mixed-sensitivity Fe4531, Fe5015, and Fe5782,
Na-sensitive Na D, and TiO$_{1}$ and TiO$_{2}$.

{\em CN$_{1}$, CN$_{2}$: } In the top left panel of Figure 3, Milky
Way GCs are systematically $\sim$0.03 mag stronger than our models in
CN$_{1}$.  In CN$_{2}$, the data tilt against our models, being weaker
at the metal-poor end but stronger at the metal-rich end compared to
our models (similar to TiO$_{2}$ in Figure 6).  The CN$_{1}$ and
CN$_{2}$ mismatch may originate from intrinsic differences between
globular cluster stars and galactic field halo stars on that our
models are based (e.g., \citealt{lan92}).  The Milky Way GC stars,
down to main sequence turn-off, are known to be generally different
from the field halo stars in the sense that the former show some
phenomena such as CH-CN bimodality (e.g., \citealt{can98};
\citealt{har03}).  As TMB03a suggested, nitrogen enhancement rather
than that of carbon might be helpful to alleviate the discrepancy
between models and the data. Also, it is noted in Tables 2 and 3 of
\citet{wfgb94} that the fitting functions for CN$_{1}$ and CN$_{2}$
are not valid below [Fe/H] = $-$1 due to lack of stars in the sample
over which one could model a fit. We are left, then, with too many
possible explanations for CN index drift.

According to TB95 and HTWB02, both CN$_{1}$ and CN$_{2}$ are heavily 
sensitive to C and N. In this study, however, as we employ the HTWB02 
treatment at the solar and super-solar metallicity regime we do not 
include C- and N-enhancement (see $\S$ 2). Hence, as depicted 
in Figure 3 both CN$_{1}$ and CN$_{2}$ (also G4300 and C$_{2}$4668) 
decrease with increasing $\alpha$-enhancement mostly due to O and Mg.

{\em G4300: } The middle left panel of Figure 3 shows that our models
of G4300 competently trace C98 data, but not for some of metal-rich
P02 GCs.  Opposite to TMB03a models, and conceivably because we do not
include C-enhancement, G4300 decreases a bit with increasing
$\alpha$-enhancement, mostly due to O. G4300 is a C- and O-sensitive
index according to TB95 and HTWB02.

{\em C$_{2}$4668:} The middle right panel of Figure 3 depicts that our
models illustriously track C98 points, while P02 GCs show a poor match as
already seen in Fig. 2 of TMB03a.  This is primarily a C-sensitive
index, but responds negatively to O-enhancement. Contrary to
TMB03a, possibly because we do not take into account C-enhancement,
C$_{2}$4668 decreases with increasing $\alpha$-enhancement mostly due
to increased O.

{\em Ca4227:} The bottom left panel of Figure 3 shows that both C98
and P02 are slightly weaker than our models at the metal-rich ([Fe/H]
$>$ $-$0.5) regime.  Ca4227 is a highly $\alpha$(Ca)-sensitive
index. In fact, it is the only Ca-sensitive Lick index according to
TB95 and HTWB02.  Without considering the effects from C- and
N-enhancement, Ca4227 significantly increases with increasing
$\alpha$-enhancement mostly due to Ca as manifested in Figure 3 at the
super-solar metallicity regime contrary to TMB03a.

{\em Ca4455:} Some notable mismatch is seen between Milky Way GCs 
and our models for Ca4455 in the bottom right panel of Figure 3.  NGC
6356 ([Fe/H] = $-$0.50) is the only common GC between C98 and P02 (C98
= 0.69 \AA, P02 = 1.48 \AA).  We suspect that P02's measurements are
maybe erroneous because it is seen in Figure 10 that C98 extensively
overlap with M31 GCs. One possibility is that Ca4455 generally
requires large systematic corrections to get from modern CCD spectra
back to the original Lick/IDS system \citep{wor04}.  Ca4455 is
misnamed in the sense that it is not an
$\alpha$- or Ca-sensitive index at all (TB95; HTWB02).  It is actually an
$\alpha$-{\em in}sensitive index as demonstrated in Figure 3.

{\em Fe4383:} From the upper left panel of Figure 4, it is seen that
Milky Way GCs are rather well traced by our models of Fe4383 though
some appear mildly weaker (by $\sim$0.5 \AA) at around [Fe/H] = $-$0.5
compared to our models.  Nevertheless, this index is a promising
substitute for the popularly used Fe5270 and Fe5335 if shorter
wavelengths are available. Fe4383 an $\alpha$-{\em anti}-sensitive
index. It decreases with increasing $\alpha$-enhancement, just as
predicted by HTWB02. According to TB95, it is the most Fe-sensitive
index (but according to HTWB02, that honor is bestowed upon Fe5335).

{\em Fe5270, Fe5335, $<$Fe$>$\footnote{$<$Fe$>$ = (Fe5270 +
Fe5335)/2.}, Fe5406, Fe5709: } The rest of Figure 4 depicts that both
C98 and P02 are fairly well tracked by our models of these
Fe-sensitive line indices.  It is particularly interesting to note
that C98's NGC 6553 and NGC 6528 (the last two most metal-rich GCs)
seem to suggest that NGC 6553, because it has slightly stronger
indices, may be slightly more metal-rich than NGC 6528, contrary to
usual literature values\footnote{\citet{bar04} present [Fe/H] =
$-$0.15 for NGC6528 and [Fe/H] = $-$0.20 for NGC6553, for instance.}.
Mainly because we are considering 
$\alpha$-enhancement at fixed [Fe/H], it is noted that 
these Fe-sensitive indices are rather $\alpha$-insensitive 
(i.e. contrary to the appearance of plots in TMB03a).

{\em Mg$_{1}$, Mg$_{2}$, Mg $\it{b}$:} It is seen from Figure 5 that
our models satisfactorily match both C98 and P02. Between NGC 6553 and
NGC 6258, NGC 6553 is again consistently stronger than NGC 6528 in
these Mg indices, according to C98.  Every Mg index is highly
sensitive to Mg abundance only, rather than $\alpha$ abundance in
general. They increase with increasing $\alpha$ mostly
due to Mg.  Mg $\it{b}$ is the most Mg-sensitive index
among them.

{\em [MgFe]\footnote{[MgFe] = $\sqrt{Mg  b \times <Fe>}$.}: }
The bottom right panel of Figure 5 depicts that our model of 
this Mg $\it{b}$, Fe5270, Fe5335 combined index successfully 
traces the Milky Way GCs throughout the entire metallicity range.

{\em Fe4531, Fe5015: } The upper right panel of Figure 6 shows that
C98 are systematically weaker (by $\sim$0.5 \AA) than our models of
Fe5015 over the entire metallicity range (opposite of CN$_{1}$, but
similar to Ca4455 and TiO$_{1}$).  These two are Ti-sensitive indices
according to TB95 and HTWB02.  Both slightly increase with increasing
$\alpha$-enhancement, mostly due to Ti.

{\em Fe5782:} In the middle left panel of Figure 6 the C98 points are
relatively stronger, with large errors, compared to our models at the
metal-rich regime.  It is not an Fe-sensitive index but it responds
negatively, albeit weakly, to increased $\alpha$.

{\em Na D: } The middle right panel of Figure 6 manifests that similar
to TMB03a, our models are generally weaker than Milky Way
GCs. Particularly, data are stronger than our models at $-$1.5 $<$
[Fe/H] $<$ 0.0. Note that the \citet{wfgb94} fitting functions were
corrected for interstellar Na D absorption, but that of the globular
cluster indices have not. TMB03a suggests that the too-strong Na D
indices may be due to the Na absorption in interstellar material, and
this is certainly true at the level of a few tenths of an \AA\ in
equivalent width. In addition, similar to CN$_{1}$ and CN$_{2}$, it is
possible that Na$-$O anti-correlation phenomenon seen in globular
cluster stars may have something to do with this mismatch (e.g.,
\citealt{gra01}).  Contrary to TMB03a, Na D heftily increases with
increasing $\alpha$-enhancement mostly due to our assumed
Na-enhancement, similar to Ca4227.

{\em TiO$_{1}$, TiO$_{2}$: } The bottom left panel of Figure 6 shows
that C98 are systematically $\sim$0.01$-$0.02 mag weaker than our
models in TiO$_{1}$. For TiO$_{2}$, we use the C98 original
measurements because \citet{bea04} did not measure this index.
Possibly because of the differences between TB95 (no molecular TiO
band effects) and HTWB02, both TiO$_{1}$ and TiO$_{2}$ increase with
increasing $\alpha$-enhancement due to Ti- and O-enhancement.

From the above comparisons, we conclude that CN$_{2}$, Ca4455, Fe5015,
Fe5782, TiO$_{2}$ may be rather ambiguous as chemical abundance
indicators.  It is also noted that some of metal line indices such as
C$_{2}$4668, Fe5335, Fe5406, Na D, TiO$_{1}$, TiO$_{2}$ do not go 
monotonically with [Fe/H] at the metal-poor end (i.e., they become 
stronger at [Fe/H] $<$ $-$1.5).  It is suggested from Fig. 7 of 
\citet{gib03} that it is of great importance to check whether the 
metallicities of Milky Way globular clusters are safely reproduced 
from the models or not before the diagnostic diagrams such as 
Mg $\it{b}$ vs. Balmer lines are used for age estimation. With some 
notable exceptions, our models match the integrated feature strengths 
of GCs with considerable accuracy.

\subsection{COMPARISON WITH M31 GLOBULAR CLUSTERS}

Having compared and examined our $\alpha$-enhanced models with Milky
Way GCs, we apply our models to M31 GCs.  In this study, we mainly use
\citet{bea04} Keck observations of M31 GCs listed in their Tables 1
and 4. Among their sample of M31 GCs, they have analyzed that 6 of
them are possibly young star clusters (YSCs: depicted as triangles in
our plots)\footnote{They are 222-277, 314-037, 321-046, 322-049,
327-053, and 380-313.}, 6 of them are intermediate-age GCs (IAGCs:
diamonds in our plots)\footnote{They are 126-184, 292-010, 301-022,
337-068, NB 16, and NB 67. Please refer to \citet{bea05} for the full
description on the IAGCs.}, and 6 are suspected as the foreground
dwarfs (small circles in our plots)\footnote{They are NB 68, NB 74, NB
81, NB 83, NB 87, and NB 91.}.  Seventeen are left as bona fide old
M31 GCs (see also \citealt{bea05}).  In our plots, however, we have
depicted the genuine old M31 GCs with high S/N ($\geq$ 60) using
bigger asterisks\footnote{They are 134-190, 158-213, 225-280, 234-290,
347-154, 365-284, 383-318, and NB 89.} compared to the ones with low
S/N (smaller asterisks)\footnote{They are 163-217, 304-028, 310-032,
313-036, 328-054, 350-162, 393-330, 398-341, and 401-344.}.  Our plots
are drawn mainly in Mg $\it{b}$ vs. Lick indices planes so that not only
could we take advantage of one of the most $\alpha$-sensitive index
(Mg $\it{b}$) but some direct comparisons with Fig. 2 of TMB03a are
possible.

Our models in Figures 7$-$9 are the same as those in Figures 1 and 2.
From Figures 7$-$9, the following is noted:
(1) in general, our models satisfactorily match M31 GCs as well as 
Milky Way GCs at 12 Gyr {\em without zero-point shifts} which 
Thomas, Maraston, \& Korn (2004; hereafter TMK04) models notably needed 
(see \citet{bea05}'s Figs. 3, 4, 5),
(2) The YSCs are consistently located around the 1 Gyr line, 
(3) The IAGCs are well-separated from the genuine 
old M31 GCs in the H$\beta$ diagram. In the H$\gamma_{A}$ diagram, they 
may overlap with old clusters. At H$\gamma_{A}$ and H$\delta_{A}$, 
the IAGCs look $\sim$3$-$5 Gyr old, 
but they appear $\sim$2$-$5 Gyr old from H$\beta$. 
The [Fe/H] of the IAGCs is confined 
between $-$1.0 and $-$0.5.

We predict that if the IAGCs are truly intermediate-age ($\sim$2$-$5
Gyr) instead of old-age ($\sim$12 Gyr) then they should NOT be
detected with the GALEX far-UV photometry (see Fig. 2 of Lee et
al. 2003).  This also has implications for other suggested
intermediate-age star clusters found among interacting galaxies (e.g.,
\citealt{gou01}).  In this respect, it is very interesting to note
that one M31 GC in particular, 311-033, has an age estimated at about 5 Gyr by
\citet{bur04} from its integrated spectra, but a recent HST CMD by
\citet{ric05} clearly shows that this globular cluster has a
well-developed blue horizontal-branch.

Figures 10$-$12 show non-Balmer indices. In all diagrams we note that
old M31 GCs mostly overlap with Milky Way GCs. Also, from the
comparison with the two most metal-rich Milky Way GCs, NGC 6553 and
NGC 6528, it seems that there is no super-solar metallicity M31 GC
among Beasley et al. sample. In fact, the most metal-rich M31 GC from
the Beasley et al. sample looks similar to the Galactic GCs NGC 6528
and NGC 6553 in Mg $\it{b}$ and it is 163-217 (small asterisk).
Overall, the metallicity range of M31 globular clusters is similar to
that of their Galactic counterparts, within sample selection effects
and small number statistics.  It is further noted that the
YSCs stand out in the moderately age-sensitive CN and G4300 diagrams.
Finally, for the purpose of weeding out foreground dwarfs (small
circles), CN$_{1}$, CN$_{2}$, Mg$_{1}$, Na D, and TiO$_{1}$ look
useful.
The detailed descriptions of some notable indices are given below.
Although we show YSCs, IAGCs, and dwarfs in our plots, we mainly
describe the genuine old M31 GCs because
our 12 Gyr models are compared with the data.

{\em CN$_{1}$, CN$_{2}$:} The top panels of Figure 10 show that M31
GCs are also systematically stronger than our models both in CN$_{1}$
and CN$_{2}$ by $\sim$0.04 mag similar to Milky Way GCs in
Fig. 3\footnote{We note that \citet{bea04} and \citet{bur04} recently
report that M31 GCs have higher nitrogen abundance compared to Milky
Way GCs using near-UV CN3883 \AA\ and NH3360 \AA\ features,
respectively.}.  {\em G4300, C$_{2}$4668: } The middle panels of
Figure 10 illustrate that M31 GCs data nicely overlap with C98 and our
models.  {\em Ca4227:} It is seen from the bottom left panel of Figure
10 that both the Milky Way and M31 GCs are systematically weaker than
our models at the metal-rich ([Fe/H] $>$ $-$0.5) regime.  {\em
Ca4455:} The bottom right panel of Figure 10 shows that M31 GCs data
overlap with C98 rather than P02 Milky Way data, albeit with some
systematic offsets from our models.

{\em Fe4383, Fe5270, Fe5335, Fe5406, Fe5709, Fe5782, Mg$_{2}$,
Fe4531:} From Figures 11 and 12, it is seen that M31 GCs generally
overlap with Milky Way GCs and with our models.  {\em Fe5015:} The
middle right panel of Figure 12 shows that M31 GCs go well with Milky
Way GCs with similar amounts of systematic offset seen in Figure 6.
{\em Na D:} The bottom right panel of Figure 12 shows that M31 GCs
generally overlap with Milky Way GCs, suggesting some similar Na
absorption in interstellar material and/or a possible Na$-$O
anti-correlation among M31 GC stars.  {\em TiO$_{1}$:} It is seen from
the bottom left panel of Figure 12 that M31 GCs mostly overlap with
C98, though with some systematic offset ($\sim$0.01$-$0.02 mag) from
our models.

\subsection{COMPARISON WITH EARLY-TYPE GALAXIES}

In this section, we investigate the $\alpha$-element enhancement 
phenomena among early-type galaxies, such as the well-known Mg enhancement 
that have been reported over the last decade (e.g., Worthey et al. 1992). 
Our primary data set is the Lick/IDS early-type galaxies of
\citet{tra98}. This data set has the advantage that it is on the
Lick/IDS index system and that it contains a large number of
galaxies. The set has two major disadvantages. First, the
observational errors for any single galaxy are large, usually too
large to have very useful age or
metallicity discrimination. Second, the additional Balmer feature
indices of \citet{wor97} are not included. We gloss over the first
problem for this paper by taking median values (really biweight
locations; see below) and we fix the second problem by publishing here
the H$\gamma$ and H$\delta$ index measurements for almost all of the
\citet{tra98} sample.

The treatment of the Balmer indices in the Lick/IDS data set proceeds
much as outlined in \citet{tra98}, and we refer readers to that paper
for a more detailed explanation. In brief, the program AUTOINDEX,
written by Jes\'us Gonz\'alez, was used to measure the index
values. This program cross-correlates spectral regions in the
immediate neighborhood of the index wavelength definition to refine
the wavelength scale before index measurement. Error estimation comes
from a ratio of the power in pixel-to-pixel (assumed mostly photon
statistics) variations to the total power in the spectrum, where the
analysis is done in Fourier-transformed spectra. This $G$ factor was
related to real measurement error by comparing with multiply-observed
galaxies and scaled appropriately, as in \citet{tra98} section 3. 

Velocity dispersion corrections were computed from artificially
broadened templates that included G and K stars and low-$\sigma$
galaxies. The $C_j(\sigma)$ multiplicative corrections are listed here
[see \citet{tra98} equation 12] as a function of velocity dispersion
$\sigma$. The \citet{tra98} Table 4 velocity dispersions were adopted.

\begin{eqnarray}
C_{\delta A}  =1 -1.111\times 10^{-5}\sigma +9.333\times 10^{-7}\sigma^2 -6.914\times 10^{-10}\sigma^3 \\
C_{\gamma A}  =1 -8.000\times 10^{-5}\sigma +4.222\times 10^{-7}\sigma^2 -1.482\times 10^{-10}\sigma^3 \\
C_{\delta F}  =1 -4.556\times 10^{-5}\sigma -7.778\times 10^{-7}\sigma^2 +9.877\times 10^{-10}\sigma^3 \\
C_{\gamma F}  =1+ 1.667\times 10^{-5}\sigma +3.778\times 10^{-7}\sigma^2 -5.926\times 10^{-10}\sigma^3
\end{eqnarray}

The velocity dispersion corrections are among the mildest of the
Lick/IDS index set. At $\sigma = 400$~km~s$^{-1}$, the corrections are
about 10\% for both H$\delta$ indices, and about 2\% for both
H$\gamma$ indices. Thus, the contribution to the total error from
velocity dispersion is almost negligible (although we included
it). The error estimation scheme may not work as well for H$\delta$ as
for H$\gamma$ because it lies at the blue end of the spectrum, and
therefore may suffer from photon noise in a somewhat more nonlinear
way than the central portion of the spectrum that was used to
calculate the Fourier $G$ parameter. Nevertheless, we present the
index measurements and errors in Table \ref{tabBalmerShort} as our
current best guess. The overall accuracy of the errors may, as in
\citet{tra98}, be as good as 5\% for the H$\gamma$ indices, but we
doubt they are quite that good for H$\delta$.

For purposes of plotting this noisy data set, the galaxy data are
binned by velocity dispersion and listed in Table 2. 
Instead of mean and standard deviation, we derive
the more robust Tukey biweight location and scale \citep{bee90,
tuk58}. We note that the smaller galaxies are observed to have a much
broader intrinsic scatter. Median measurement error is
about the same for the smaller galaxies as for the larger ones, so the
increased scatter in small galaxies is intrinsic to the galaxies, not
an artifact of photon statistics. The implication is that the driving
parameters of the integrated spectrum (age, metallicity, nebular
emission) are more varied in the small galaxies compared to the large
ones. This agrees with earlier inferences from the same Lick/IDS data set
\citep{wor96,wor98} but also from considerations of the Mg-$\sigma$
diagram \citep{ben93,wor03}, where nebular emission is not a factor.

In Figures 13$-$19, the $\sigma$-binned median galaxy indices are
depicted as large open squares. In addition to the median data, two
individual galaxies with low observational errors, M31 (big open
circle on the metal-rich side) and M32 (big open circle on the
metal-poor side), are also plotted. Figures 13 and 14 show Balmer
index diagrams, while Figures 15$-$19 show metallic features. What 
we look into from these plots are the SSP-equivalent ages and 
metallicities of galaxies because they are generally composed of 
mixtures of stellar populations with a range of ages and metallicities.

Figure 13 shows a comparison of Lick/IDS early-type galaxies
with our models in Mg $\it{b}$ vs. H$\beta$
(left panel) and in Fe5270 vs. H$\beta$ (right panel). 
M31 star clusters of all ages are also plotted. 
The top pair of panels has [$\alpha$/Fe] = 0, the middle pair of
panels has [$\alpha$/Fe] = +0.3, and the bottom pair of panels has
[$\alpha$/Fe] = +0.6, applied only at [Fe/H] = 0.0 and +0.5 as usual.
The high $\alpha$-sensitivity of Mg $\it{b}$ (left panel)
and the $\alpha$-$\it{in}$sensitivity of Fe5270 (right panel) are
abundantly illustrated.  Figure 14 is a clone of Figure 13, except using
H$\gamma_{A}$ instead of H$\beta$ in order to show the
$\alpha$-sensitivity of H$\gamma_{A}$\footnote{All our other models of
high-order Balmer line indices (H$\gamma_{F}$, H$\delta_{A}$, and
H$\delta_{F}$) are also similarly significantly $\alpha$-sensitive
confirming the previous work of TMK04 (cf., \citealt{tan04}).
They are available at our website, http://astro.wsu.edu/hclee/sp.html.}.

As we apply our various $\alpha$-enhanced models to the data in
Figures 13 and 14, we find that vastly different age, [Fe/H], and
[$\alpha$/Fe] estimation is possible for the early-type galaxies
depending upon the selected combination of metal lines and Balmer
lines.  The well-known Mg enhancement is prominent from upper panels
of Figures 13 and 14 as galaxy data are compared with our models
from solar-scaled isochrones without HTWB02 treatment. It is seen from
middle panels of Figures 13 and 14, however, that our models of
[$\alpha$/Fe] = +0.3 dex consistently give a similar estimation of age
and metallicity both from Mg $\it{b}$ and Fe5270 for high-$\sigma$
galaxies.

The low $\sigma$ galaxies ($\sigma$ $<$ 100 km/sec) that have the
lowest values in Mg $\it{b}$ and Fe5270 have consistent ages only in
the [$\alpha$/Fe]=0 panels, as one would expect from, e.g. Figure 17.
The median low $\sigma$ galaxy also looks younger than high $\sigma$
counterparts in Figures 13 and 14 (see also \citealt{cal03}), but of
course the scatter in this bin is larger, so old-and-small galaxies do
exist.

We turn now to the metallic indices plotted in Figures 15$-$19.
The Lick IDS early-type galaxies plotted against our models, particularly
from Figure 17, show that they are generally of $-$0.5 $\leq$ [Fe/H]
$\leq$ +0.2 with [$\alpha$/Fe] $\sim$ +0.3 dex.  
It is also seen from Figure 17 that the [$\alpha$/Fe] of galaxies, at
least as traced by Mg and Na, seems 
to increase with increasing velocity dispersion \citep{tra00b}. 
As Milky Way and bona fide old M31 GCs are compared with Lick/IDS early-type 
galaxies in Figures 15$-$19, the following details are also noted.

{\em Mg:} The well-known Mg enhancement phenomenon of early-type
galaxies is again confirmed from our models of Mg $\it{b}$ vs. Fe4383
in the top left panel of Figure 17 show this issue more clearly,
because Mg $\it{b}$ is strongly $\alpha$-sensitive while Fe4383 is
$\alpha$-$\it{anti}$-sensitive.

{\em CN$_{1}$, CN$_{2}$:} The top panels of Figure 15 show that
Lick/IDS early-type galaxies are systematically of lower values than
the Milky Way and M31 GCs both in CN$_{1}$ and CN$_{2}$ by
$\sim$0.03$-$0.04 mag (see also Fig. 3 of \citealt{bea04}). Assuming
no problems with the data, this points out a distinct difference
between globular clusters and early-type galaxies.  We speculate that
the globular cluster environment is perhaps special and different from
that of galaxies, and that this is manifested in the CN index
strengths either due to N abundance effects, or perhaps a CH-CN
bimodality phenomenon. The same behavior is seen in CN diagrams versus
Fe indices.

{\em Ca4227:}
As shown in the bottom left panel of Figure 15, all the metal-rich
([Fe/H] $>$ $-$0.5) Milky Way and M31 GCs and Lick/IDS early-type
galaxies are of systematically lower Ca4227 values compared to our
models.  The ``Ca underabundance'' (e.g., TMB03b; \citealt{cen04}) of 
early-type galaxies is
manifested here. Figure 16 shows, however, that Ca4227 vs. Fe
indices does {\em not} show any significant under- or
over-abundance. That is, in diagrams involving Fe indices, the
galaxies follow the scaled-solar models. We tend to agree with
\citet{wor98} that while Mg, Na, and perhaps N are enhanced with
respect to Fe, Ca is not.

{\em TiO$_{1}$ and TiO$_{2}$:}
It is seen from the bottom left panel of Figure 19 that Lick/IDS
early-type galaxies overlap and extend the metal-rich end of the Milky
Way (C98) and M31 GCs. Curiously, many GCs appear to have much stronger TiO
absorption than the most metal-rich of galaxies. Perhaps some stochastic
effect can account for this, but our favorite explanation is another
appeal to the slippery transformation from CCD to Lick/IDS systems. In
this case, the TiO indices are both broad in wavelength and at the red 
end of the spectrum, both of which leave the index especially sensitive
to systematic errors induced by spectrophotometric shape effects.

\section{SUMMARY \& CONCLUSION}

In this study we presented new simple stellar population models that
predict 25 Lick IDS index strengths and cover a broad range of age and
metallicity. They include $\alpha$-element enhancement by virtue of
the fact that they are based on the $\alpha$-enhanced $Y^{2}$
isochrones and HTWB02 response functions for Lick IDS indices. 
The effects from horizontal-branch stars are fully
incorporated. 

Among Balmer lines, a significant $\alpha$-enhancement effect was
noted for H$\gamma$ and H$\delta$, but H$\beta$ is rather
$\alpha$-$\it{in}$sensitive. We also find that our 5 Gyr and 12 Gyr
models of H$\gamma$ and H$\delta$ overlap at the metal-poor regime
because of blue horizontal-branch stars.  It is further noted that
younger than 1 Gyr and for metal-poor populations, Balmer lines become
weaker because the main-sequence turnoff becomes hotter than 10,000 K.

Our $\alpha$-enhanced models are presented at fixed [Fe/H],
therefore making it straightforward to test our models using Milky
Way globular clusters that have independently-estimated mean
[Fe/H] and [$\alpha$/Fe].  At the solar and super-solar metallicity
end, we have further employed the updated Lick IDS index response functions by
HTWB02 so that we could assess element enhancement
in early-type galaxies. Our models should be
useful for distant unresolved stellar systems in order to extract some
useful information on their age, metallicity, and $\alpha$-enhancement.

From the comparisons of our models with observations of Milky Way and
M31 GCs, we find that our models of Balmer lines satisfactorily trace
them at 12 Gyr {\em without the need for zero-point shifts} which some
previous models require.  We also verify the plausible existence of
young and intermediate-age star clusters in M31 that Beasley et
al. (2004, 2005) recently claimed.  We derive their ages at around
$\sim$1 Gyr and $\sim$2$-$5 Gyr, respectively, from our models. We
predict that these latter ones are truly intermediate-age ($\sim$2$-$5
Gyr) instead of old-age ($\sim$12 Gyr) if they are {\em not} detected
with GALEX far-UV photometry.  We do not find any super-solar
metallicity M31 GCs from \citet{bea04}'s sample. The metallicity range of
M31 GCs in \citet{bea04} is similar to that of their Galactic
counterparts.  Contrary to the literature values, the Milky Way GC NGC
6553 appears slightly more metal-rich than NGC 6528 based on the
strength of their metal indices.

We confirm and understand the well-known Mg enhancement phenomenon
among metal-rich early-type galaxies.  Our models with [$\alpha$/Fe] =
+0.3 dex consistently provide similar estimation of age and
metallicity both from Mg and Fe indices.  Also, it is noted that
[$\alpha$/Fe] of galaxies, as traced by Mg and Na indices, seems to
increase with increasing velocity dispersion (but Ca tracks
Fe). Moreover, and independent of models, GCs and galaxies follow
different tracks in diagrams involving CN$_{1}$ and CN$_{2}$. This may
be telling us something about N abundance or CH-CN bimodality that is
generically different in the globular cluster environment as opposed
to the elliptical galaxy environment. It is tempting to speculate that
either the extremely high stellar density in globular clusters or some
type of partial self-enrichment in globular clusters is the physical
driver behind this intriguing difference.  Finally, we stress that it
is of considerable importance to use these $\alpha$-enhanced Lick indices 
in order to properly assess the mean age and chemical compositions of
early-type galaxies.

\acknowledgments

It is a great pleasure to thank Brad Gibson, Michael Beasley, Rob
Proctor, Ricardo Schiavon, Daniel Thomas, Claudia Maraston, and Markus
Kissler-Patig for many helpful discussions.  We thank Sukyoung Yi for
providing us the mass loss table for the $\alpha$-enhanced $Y^{2}$
isochrones. We also thank an anonymous referee for her/his perspicacious 
reading of the manuscript which improved this paper significantly.
Support for this work was provided by the NSF \# 0307487.


\begin{deluxetable}{lrrrrrrrr}
\tabletypesize{\small}
\tablecaption{Balmer Indices for Lick/IDS Galaxies \label{tabBalmerShort}}
\tablehead{\colhead{Object} & 
\colhead{H$\delta_A$} & \colhead{$\sigma_{H\delta A}$ } &
\colhead{H$\gamma_A$} & \colhead{$\sigma_{H\gamma A}$ } &
\colhead{H$\delta_F$} & \colhead{$\sigma_{H\delta F}$ } &
\colhead{H$\gamma_F$} & \colhead{$\sigma_{H\gamma F}$ } }

\startdata
A 569A   &    -2.596&  1.322& -4.961&  0.810& -0.178&  0.508& -1.326&  0.460\\
IC 171   &    -2.017&  2.215& -5.731&  1.371& -0.506&  0.874& -1.123&  0.772\\
IC 179   &    -3.214&  1.704& -5.654&  1.046& -1.263&  0.669& -1.665&  0.593\\
IC 310   &    -0.176&  1.355& -3.612&  0.830&  0.779&  0.525& -1.324&  0.472\\
IC 1696  &    -1.466&  1.251& -5.260&  0.776&  0.945&  0.499& -0.777&  0.437\\
IC 1907  &    -2.842&  1.972& -5.305&  1.206& -0.987&  0.760& -1.048&  0.680\\
IC 2955  &    -2.277&  1.372& -7.964&  0.848&  0.064&  0.538& -2.808&  0.488\\
IC 4051  &    -2.311&  2.604& -4.136&  1.598& -0.010&  1.003& -0.813&  0.900\\
NGC 80   &    -3.224&  2.006& -5.068&  1.209& -0.293&  0.739& -1.626&  0.687\\
NGC 83   &    -1.130&  2.055& -3.526&  1.253& -0.067&  0.778& -1.916&  0.713\\
NGC 128  &   \nodata&\nodata& -5.698&  0.888&  0.010&  0.562& -1.470&  0.503\\
NGC 194  &    -1.074&  1.108& -4.709&  0.682&  0.154&  0.431& -0.923&  0.386\\
NGC 205  &     6.707&  1.178&  4.951&  0.739&  4.728&  0.498&  4.621&  0.419\\
NGC 221  &    -1.138&  0.264& -4.341&  0.166&  0.669&  0.109& -0.416&  0.093\\
NGC 224  &    -2.514&  0.276& -6.405&  0.167& -0.027&  0.106& -1.394&  0.107\\
NGC 227  &    -0.854&  1.112& -6.562&  0.677&  0.715&  0.427& -2.244&  0.403\\
NGC 315  &    -2.909&  2.305& -6.544&  1.375&  1.080&  0.841& -2.967&  0.802\\

\enddata

\tablecomments{Table \ref{tabBalmerShort} appears in its entirety in
the electronic edition of the Astrophysical Journal Supplement Series.}

\end{deluxetable}



\begin{deluxetable}{rrrrrrrrrrrrrr}
\tablewidth{570pt}
\rotate
\tabletypesize{\tiny}
\tablenum{2.a}
\tablecaption{Biweight Locations and Scales for the Lick/IDS Galaxies, Binned by Velocity Dispersion\tablenotemark{a} \label{tabMedians}}

\tablehead{ \colhead{LOSVD km s$^{-1}$} &
\colhead{CN$_1$} &
\colhead{CN$_2$} &
\colhead{Ca4227} &
\colhead{G4300} &
\colhead{Fe4383} &
\colhead{Ca4455} &
\colhead{Fe4531} &
\colhead{C$_2$4668} &
\colhead{H$\beta$} &
\colhead{Fe5015} &
\colhead{Mg$_1$} &
\colhead{Mg$_2$} &
\colhead{Mg $b$} }

\startdata
     $< 100$     & -0.022  &   0.030 &   0.880 &   4.208 &   3.905 &   1.375 &   2.991 &   5.089 &   2.204 &   4.852 &   0.068 &   0.196 &   3.021 \\
                            &   0.071 &   0.063 &   0.505 &   2.080 &   1.704 &   0.495 &   0.936 &   2.552 &   0.802 &   1.087 &   0.043 &   0.067 &   1.094 \\
                            &  29     & 29      & 29      & 27      & 29      & 29      & 29      & 29      & 26      & 25      & 27      & 27      & 29      \\
   100 --- 150   &  0.047  &   0.082 &   1.024 &   5.096 &   4.733 &   1.576 &   3.435 &   6.481 &   1.687 &   4.975 &   0.112 &   0.255 &   4.012 \\
                            &   0.048 &   0.048 &   0.308 &   0.941 &   0.999 &   0.333 &   0.464 &   1.400 &   0.561 &   0.872 &   0.028 &   0.042 &   0.523 \\
                            &  47     & 47      & 47      & 45      & 47      & 47      & 47      & 47      & 46      & 44      & 47      & 47      & 47      \\
   150 --- 200   &  0.090  &   0.125 &   1.207 &   5.437 &   5.121 &   1.711 &   3.644 &   7.449 &   1.585 &   5.532 &   0.140 &   0.290 &   4.465 \\
                            &   0.042 &   0.047 &   0.449 &   0.924 &   1.081 &   0.410 &   0.675 &   1.463 &   0.463 &   0.712 &   0.025 &   0.033 &   0.575 \\
                            &  86     & 86      & 86      & 83      & 86      & 86      & 86      & 86      & 82      & 80      & 86      & 86      & 85      \\
   200 --- 240   &  0.106  &   0.147 &   1.312 &   5.548 &   5.298 &   1.742 &   3.674 &   7.871 &   1.427 &   5.415 &   0.152 &   0.307 &   4.799 \\
                            &   0.033 &   0.037 &   0.437 &   0.854 &   0.845 &   0.362 &   0.681 &   1.176 &   0.446 &   1.038 &   0.019 &   0.027 &   0.477 \\
                            &  80     & 80      & 81      & 78      & 81      & 81      & 81      & 81      & 80      & 79      & 80      & 80      & 80      \\
   240 --- 280   &  0.110  &   0.151 &   1.395 &   5.456 &   5.244 &   1.781 &   3.672 &   8.078 &   1.420 &   5.451 &   0.160 &   0.317 &   4.963 \\
                            &   0.035 &   0.040 &   0.366 &   0.867 &   0.916 &   0.372 &   0.614 &   1.177 &   0.414 &   1.226 &   0.018 &   0.022 &   0.409 \\
                            &  71     & 71      & 68      & 71      & 71      & 71      & 71      & 71      & 69      & 68      & 70      & 70      & 70      \\
   $>280$        &  0.134  &   0.173 &   1.374 &   5.388 &   5.297 &   1.846 &   3.915 &   8.621 &   1.393 &   5.550 &   0.169 &   0.332 &   5.127 \\
                            &   0.027 &   0.031 &   0.500 &   0.899 &   1.013 &   0.489 &   0.727 &   1.314 &   0.540 &   1.061 &   0.017 &   0.021 &   0.392 \\
                            &  42     & 42      & 42      & 42      & 42      & 42      & 42      & 42      & 41      & 41      & 41      & 41      & 42      \\

\enddata
\end{deluxetable}

\begin{deluxetable}{rrrrrrrrrrrrrr}
\tablewidth{576pt}
\rotate
\tabletypesize{\tiny}
\tablenum{2.b}
\tablecaption{}

\tablehead{\colhead{LOSVD km s$^{-1}$} &
\colhead{Fe5270} &
\colhead{Fe5335} &
\colhead{Fe5406} &
\colhead{Fe5709} &
\colhead{Fe5782} &
\colhead{Na D} &
\colhead{TiO$_1$} &
\colhead{TiO$_2$} &
\colhead{H$\delta_A$} &
\colhead{H$\gamma_A$} &
\colhead{H$\delta_F$} &
\colhead{H$\gamma_F$} 
}

\startdata
 $< 100$   &   2.517 &   2.187 &   1.485 &   0.935 &   0.682 &   2.952 &   0.034 &   0.059 &   0.698 &  -3.227 &   1.481 &  -0.436 & \\
 &   0.641 &   0.601 &   0.492 &   0.423 &   0.300 &   0.790 &   0.011 &   0.019 &   3.263 &   4.369 &   1.669 &   2.436 & \\
 & 29      & 29      & 27      & 25      & 24      & 28      & 29      & 20      & 29      & 27      & 29      & 29      & \\
100 --- 150  &   2.863 &   2.452 &   1.716 &   0.992 &   0.794 &   3.579 &   0.038 &   0.079 &  -1.205 &  -5.260 &   0.450 &  -1.516 & \\
 &   0.437 &   0.418 &   0.313 &   0.228 &   0.279 &   0.792 &   0.011 &   0.016 &   1.553 &   1.894 &   0.757 &   1.176 & \\
 & 47      & 47      & 45      & 45      & 45      & 47      & 47      & 28      & 45      & 45      & 46      & 46      & \\
150 --- 200  &   3.045 &   2.785 &   1.925 &   1.009 &   0.852 &   4.575 &   0.039 &   0.096 &  -2.153 &  -6.103 &   0.206 &  -1.917 & \\
 &   0.384 &   0.426 &   0.371 &   0.236 &   0.320 &   0.812 &   0.013 &   0.013 &   1.129 &   1.178 &   0.714 &   0.760 & \\
 & 85      & 83      & 76      & 71      & 81      & 86      & 86      & 44      & 82      & 80      & 85      & 86      & \\
200 --- 240  &   3.107 &   2.795 &   1.945 &   0.977 &   0.862 &   4.942 &   0.041 &   0.100 &  -2.344 &  -6.154 &  -0.017 &  -1.951 & \\
 &   0.375 &   0.440 &   0.398 &   0.280 &   0.335 &   0.760 &   0.011 &   0.010 &   1.105 &   1.260 &   0.640 &   0.780 & \\
 & 79      & 79      & 76      & 69      & 79      & 81      & 81      & 30      & 80      & 75      & 80      & 81      & \\
 240 --- 280  &   3.018 &   2.687 &   1.983 &   0.965 &   0.827 &   5.192 &   0.045 &   0.098 &  -2.359 &  -6.317 &   0.013 &  -1.952 & \\
 &   0.417 &   0.475 &   0.367 &   0.302 &   0.258 &   0.703 &   0.011 &   0.014 &   1.057 &   1.076 &   0.518 &   0.566 & \\
 & 70      & 66      & 67      & 64      & 70      & 71      & 71      & 19      & 70      & 70      & 70      & 70      & \\
$>280$   &   3.230 &   2.712 &   1.940 &   0.883 &   0.911 &   5.718 &   0.045 &   0.095 &  -2.860 &  -6.413 &  -0.146 &  -2.112 & \\
 &   0.557 &   0.580 &   0.422 &   0.333 &   0.424 &   0.582 &   0.013 &   0.016 &   0.763 &   0.866 &   0.557 &   0.472 & \\
 & 42      & 41      & 39      & 36      & 40      & 42      & 42      &  8      & 41      & 42      & 42      & 42      & \\
\enddata

\tablenotetext{a}{Table is arranged in trios of rows. The first row is the
biweight location (median), the second is the biweight scale, a measure of the
scatter that is equivalent to a Gaussian $\sigma$ for Gaussian-distributed
data, and the third is the number of galaxies used in the biweight computation
for each velocity dispersion bin and index. ``LOSVD'' is line of sight
velocity dispersion.}

\end{deluxetable}



\clearpage

\begin{figure}
\epsscale{.8}
\plotone{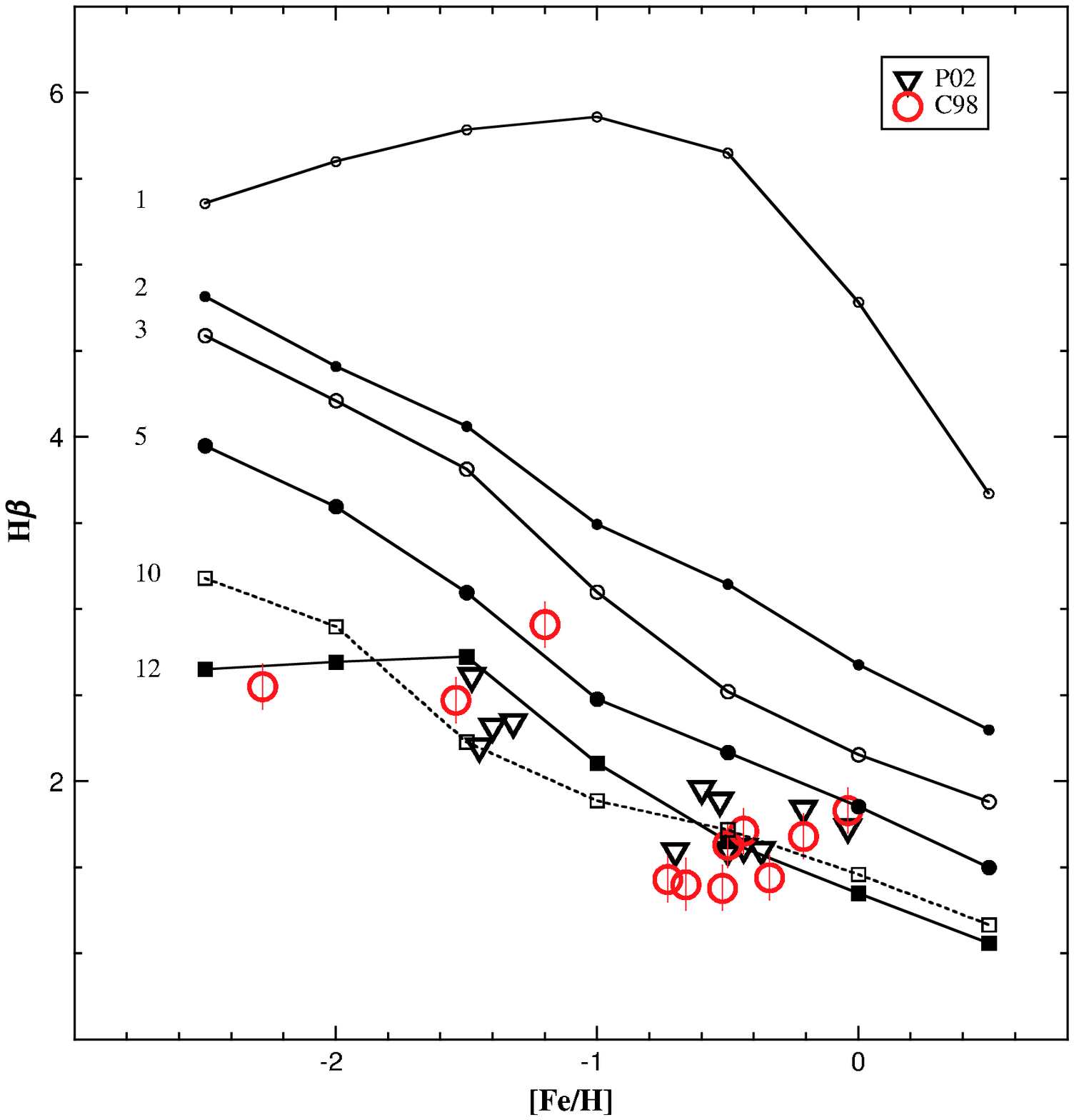}
\caption{Models of H$\beta$ are compared with Milky Way GCs. Model
ages (in Gyr) are indicated on the left. Model points are shown at
[Fe/H] = $-$2.5, $-$2.0, $-$1.5, $-$1.0, $-$0.5, 0.0, and +0.5.  At
[Fe/H] = 0.0 and +0.5, models that incorporate index response
functions with [$\alpha$/Fe] = +0.3 dex are shown.  Cohen et
al. (1998; C98, {\em circles}) and Puzia et al. (2002; P02, {\em
triangles}) observations are overlaid. The [Fe/H] of Milky Way GCs are
from Harris (1996).  Note that our 12 Gyr models trace the Milky Way
GCs without zero-point shifts (see text).  The 12 Gyr and 10 Gyr
lines cross due to HB morphology effects.}
\end{figure}

\begin{figure}
\epsscale{1.0}
\plotone{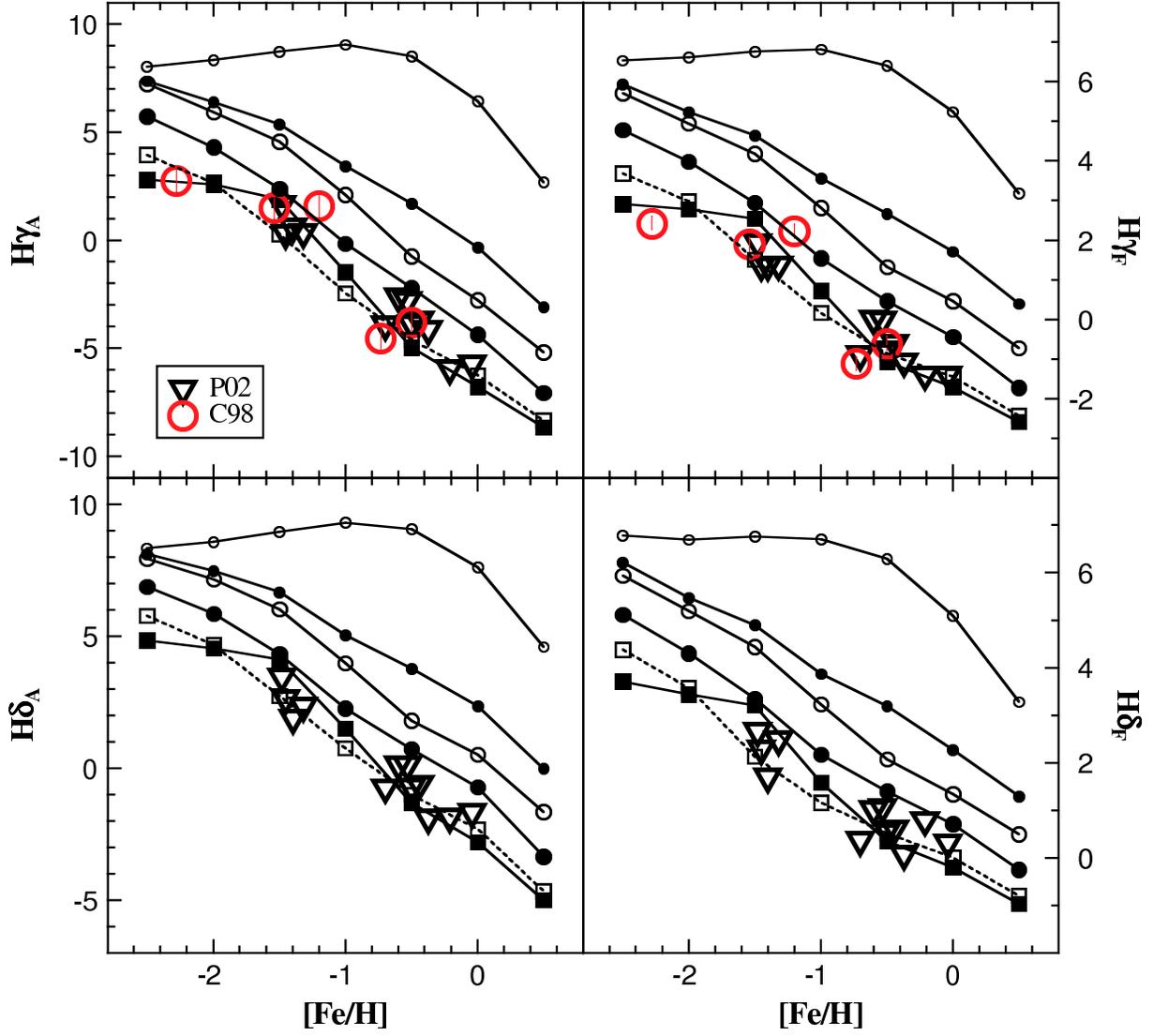}
\caption{Same as Figure 1, but our models of H$\gamma$ and H$\delta$
are compared with Milky Way GCs (see text). Symbols 
for our models are the same as Figure 1.}
\end{figure}

\begin{figure}
\epsscale{0.8}
\plotone{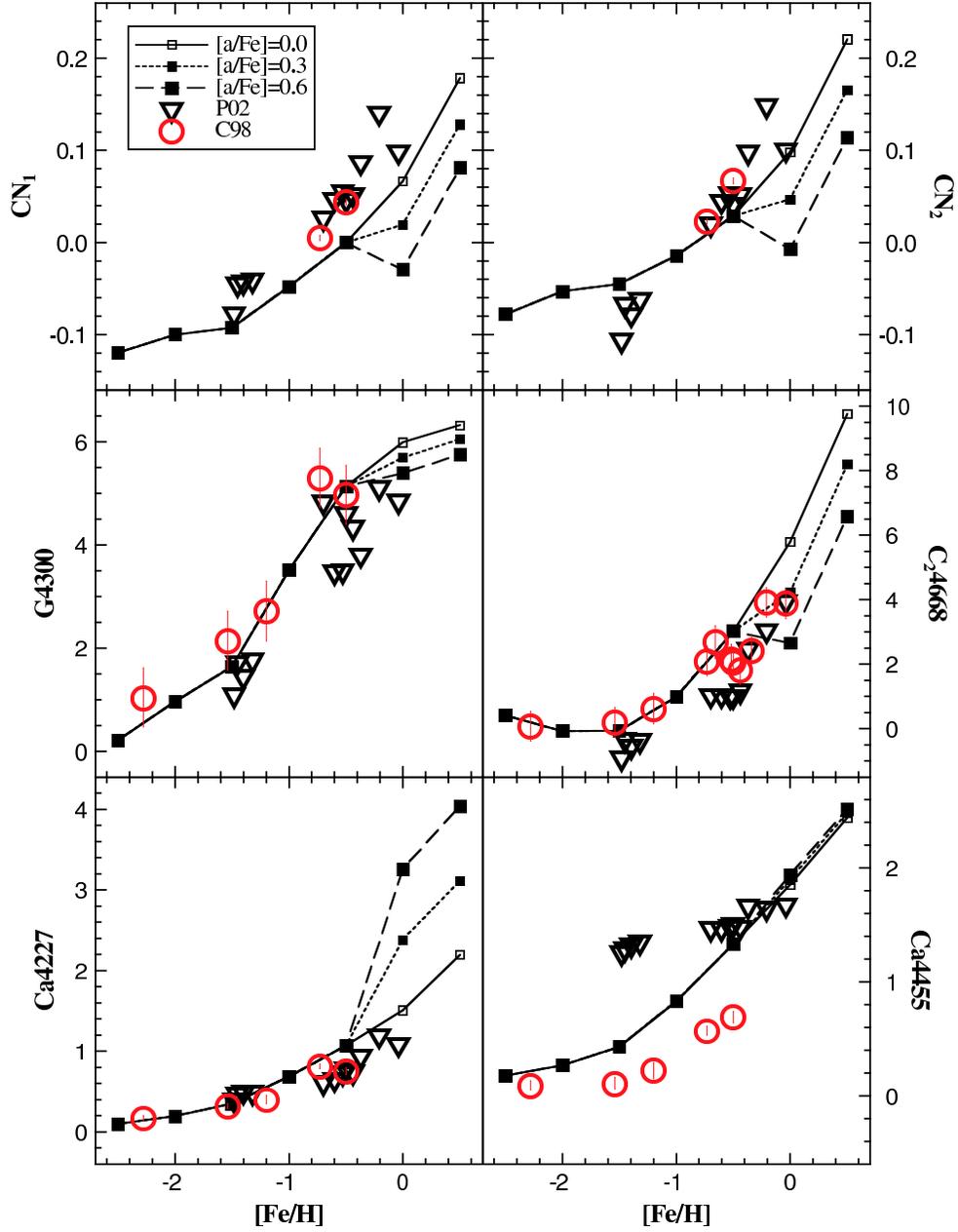}
\caption{12 Gyr models of CN$_{1}$, CN$_{2}$, G4300, C$_{2}$4668,
Ca4227, and Ca4455 are compared with Milky Way GCs. For models of
[Fe/H] = $-$2.5, $-$2.0, $-$1.5, and $-$1.0, $Y^{2}$ isochrones with
[$\alpha$/Fe] = +0.3 dex are used. For [Fe/H] = $-$0.5, $Y^{2}$
isochrones with [$\alpha$/Fe] = +0.15 dex are employed.  For those of
[Fe/H] = 0.0 and +0.5, solar-scaled $Y^{2}$ isochrones are used ({\em
solid lines}) and HTWB02 response functions with [$\alpha$/Fe] = +0.3
dex ({\em dotted lines}) and +0.6 dex ({\em dashed lines}) are applied.}
\end{figure}

\begin{figure}
\epsscale{0.8}
\plotone{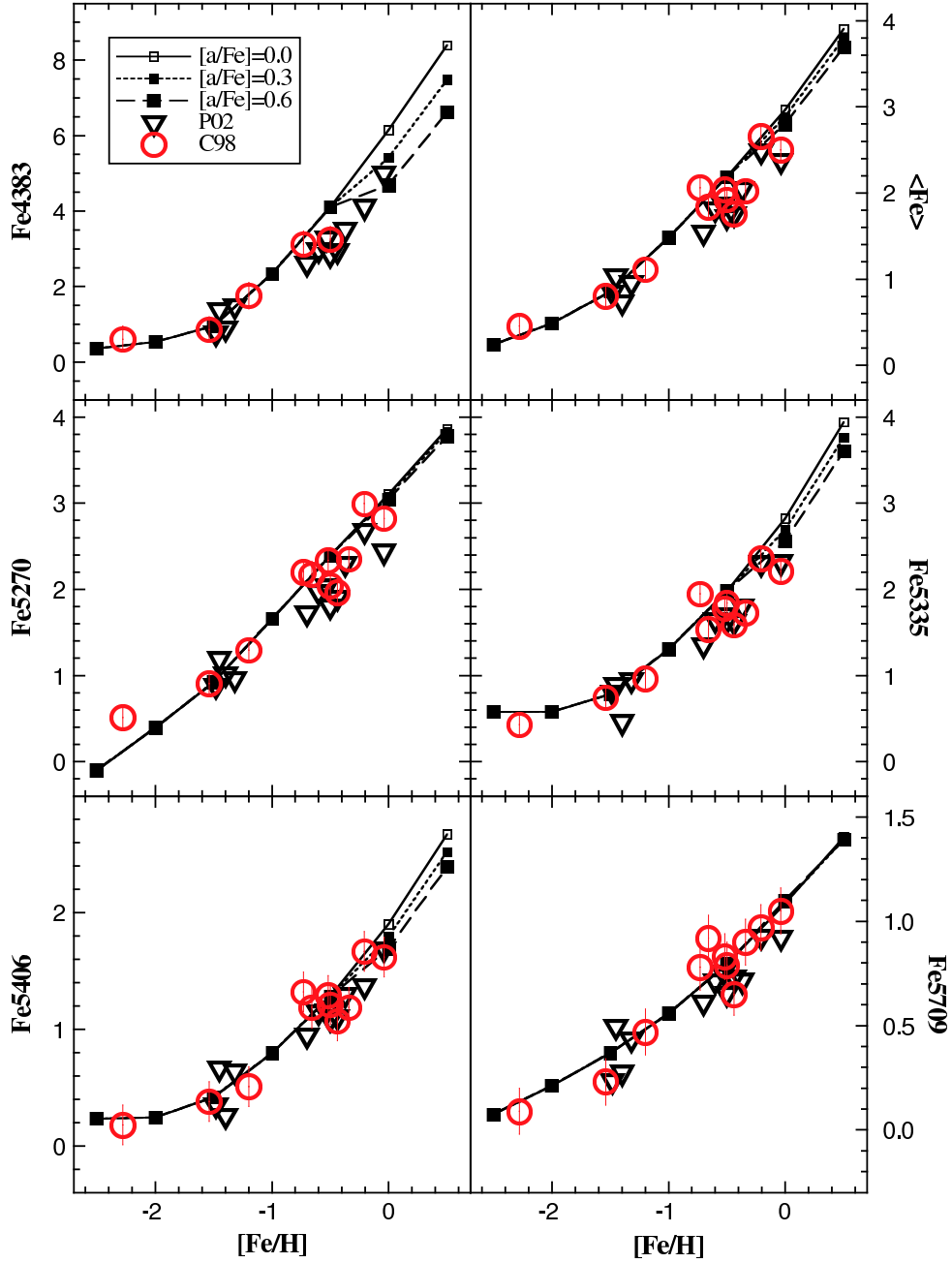}
\caption{Same as Figure 3, but for different indices. 12 Gyr models of 
Fe4383, $<$Fe$>$, Fe5270, Fe5335, Fe5406, and Fe5709 are compared with 
Milky Way GCs. Symbols are the same as Figure 3.}
\end{figure}

\clearpage

\begin{figure}
\epsscale{1.0}
\plotone{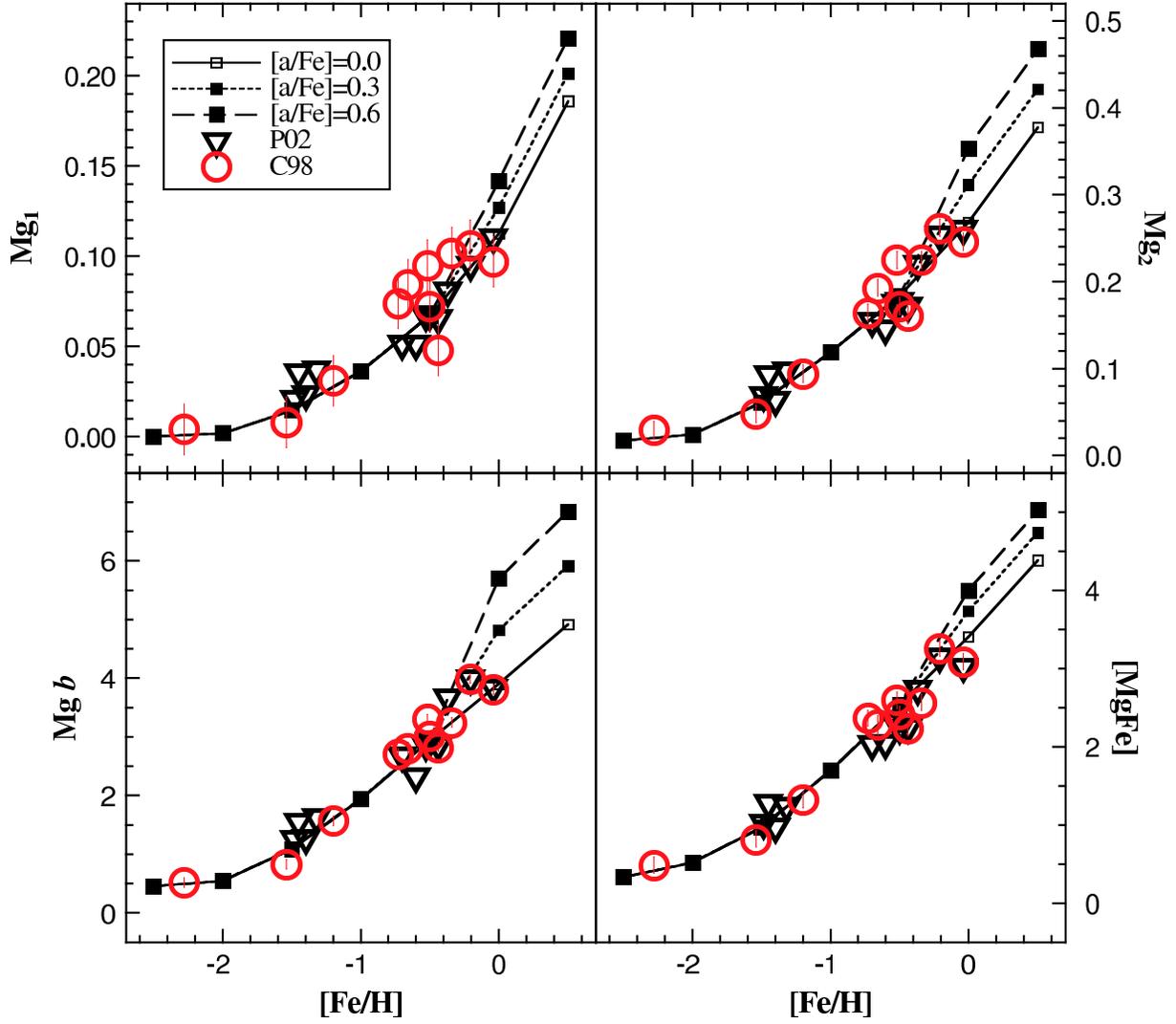}
\caption{Same as Figure 3, but for different indices. 12 Gyr models of 
Mg$_{1}$, Mg$_{2}$, Mg $\it{b}$, and [MgFe] are compared with 
Milky Way GCs. Symbols are the same as Figure 3.}
\end{figure}

\begin{figure}
\epsscale{0.8}
\plotone{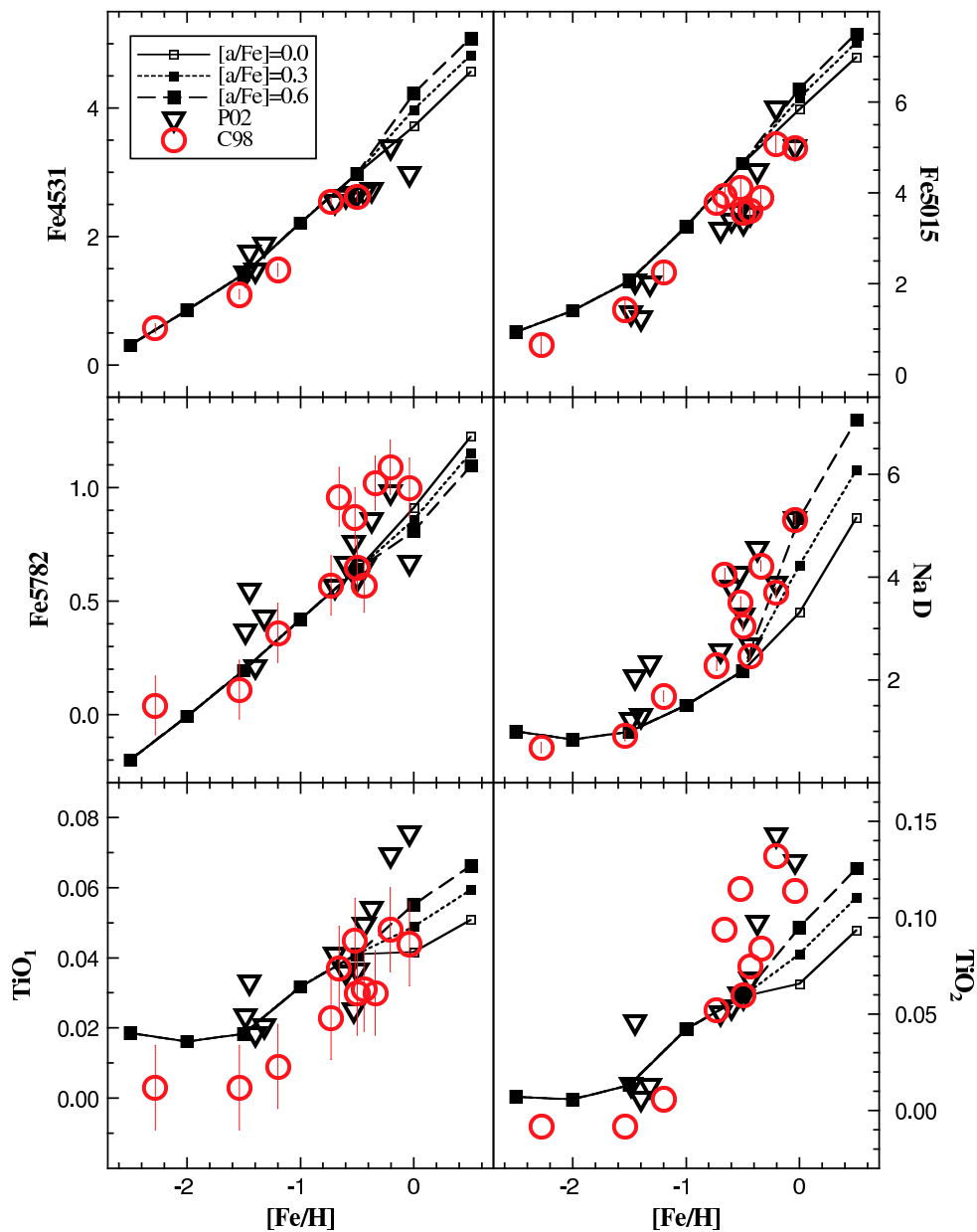}
\caption{Same as Figure 3, but for different indices. 12 Gyr models of 
Fe4531, Fe5015, Fe5782, Na D, TiO$_{1}$, and TiO$_{2}$ are compared with 
Milky Way GCs. Symbols are the same as Figure 3.}
\end{figure}

\begin{figure}
\epsscale{0.9}
\plotone{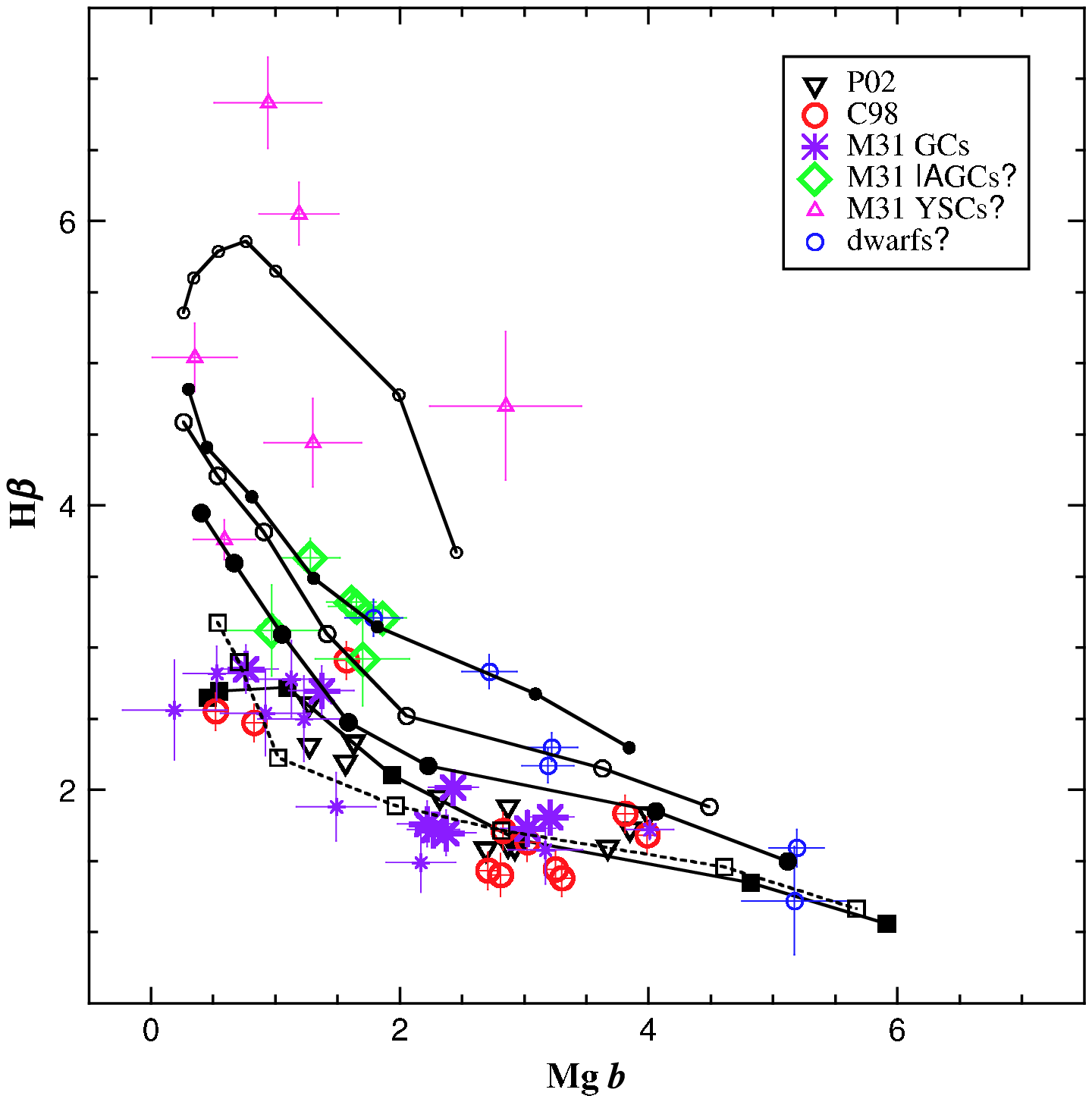}
\caption{Beasley et al. (2004) M31 star clusters are compared with
Milky Way GCs and with our models in Mg $\it{b}$ vs. H$\beta$.
Symbols for our models and Milky Way GCs are the same as Figure 1.
Bona fide M31 GCs ({\em asterisks}), bona fide M31 GCs with S/N $\geq$
60 ({\em bigger asterisks}), suspected IAGCs ({\em diamonds}), YSCs
({\em triangles}), and foreground dwarf stars ({\em small circles})
are also plotted. }
\end{figure}

\begin{figure}
\epsscale{0.9}
\plotone{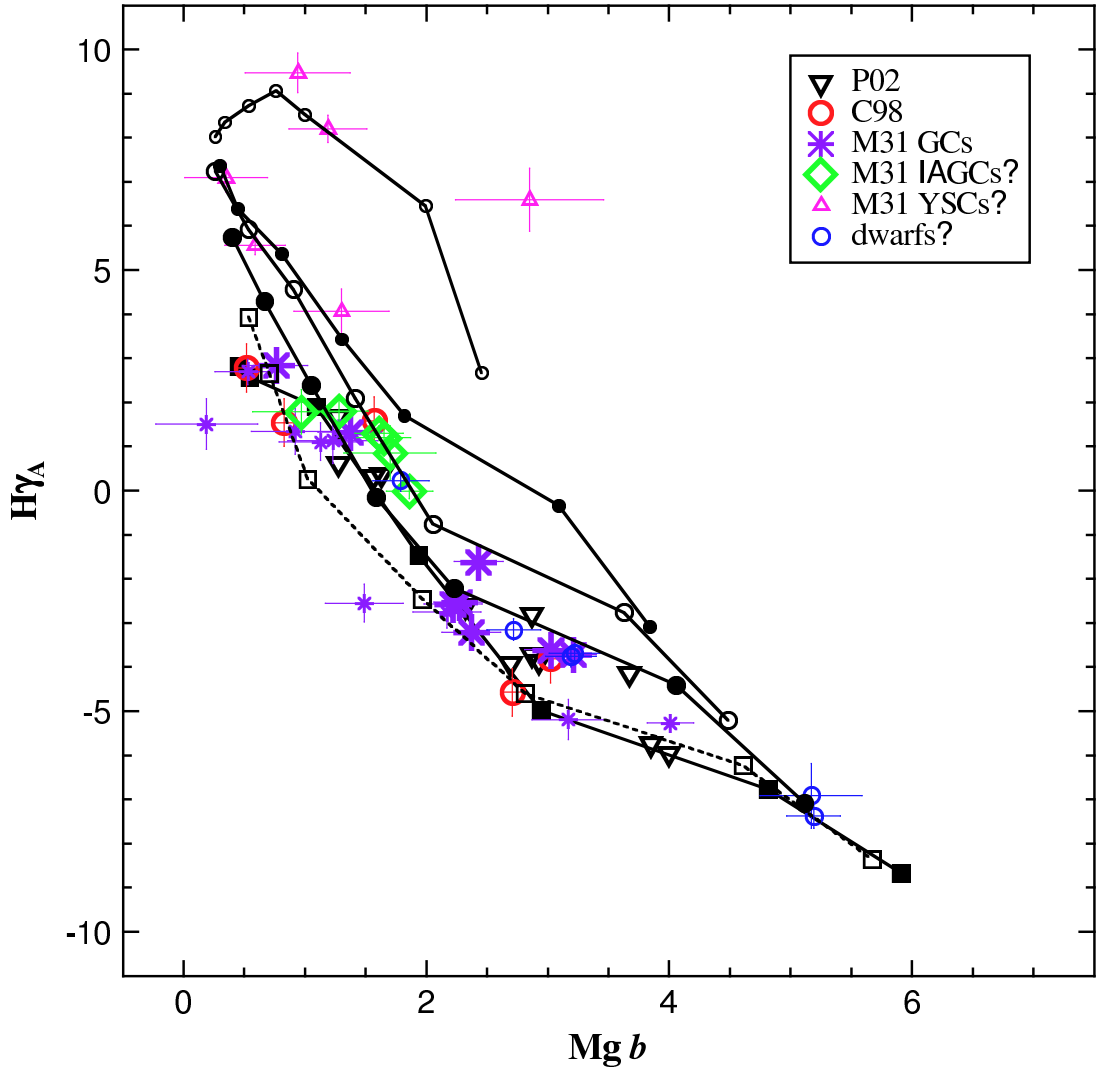}
\caption{Same as Figure 7, but with Mg $\it{b}$ vs. H$\gamma_{A}$.}
\end{figure}

\clearpage

\begin{figure}
\epsscale{0.9}
\plotone{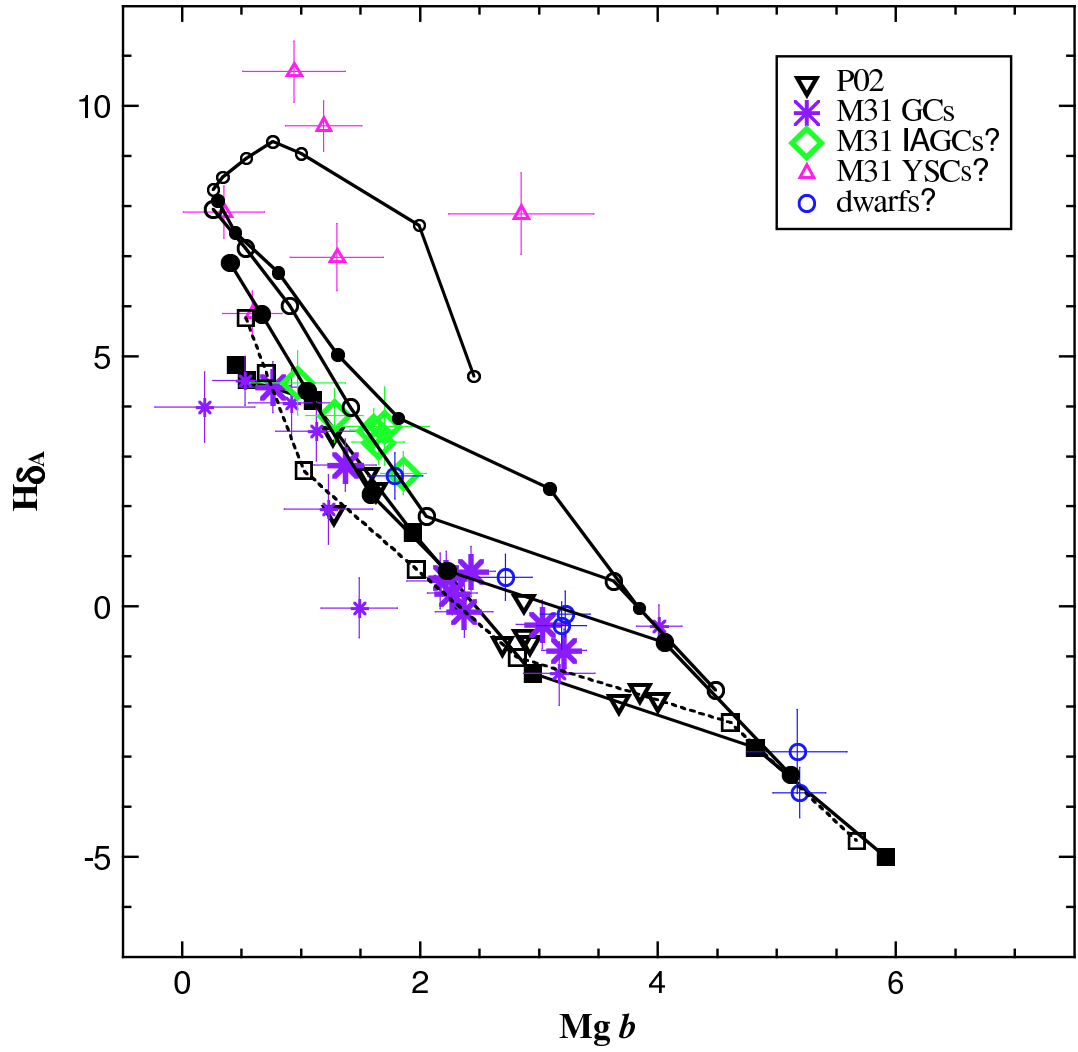}
\caption{Same as Figure 7, but with Mg $\it{b}$ vs. H$\delta_{A}$.}
\end{figure}

\begin{figure}
\epsscale{0.8}
\plotone{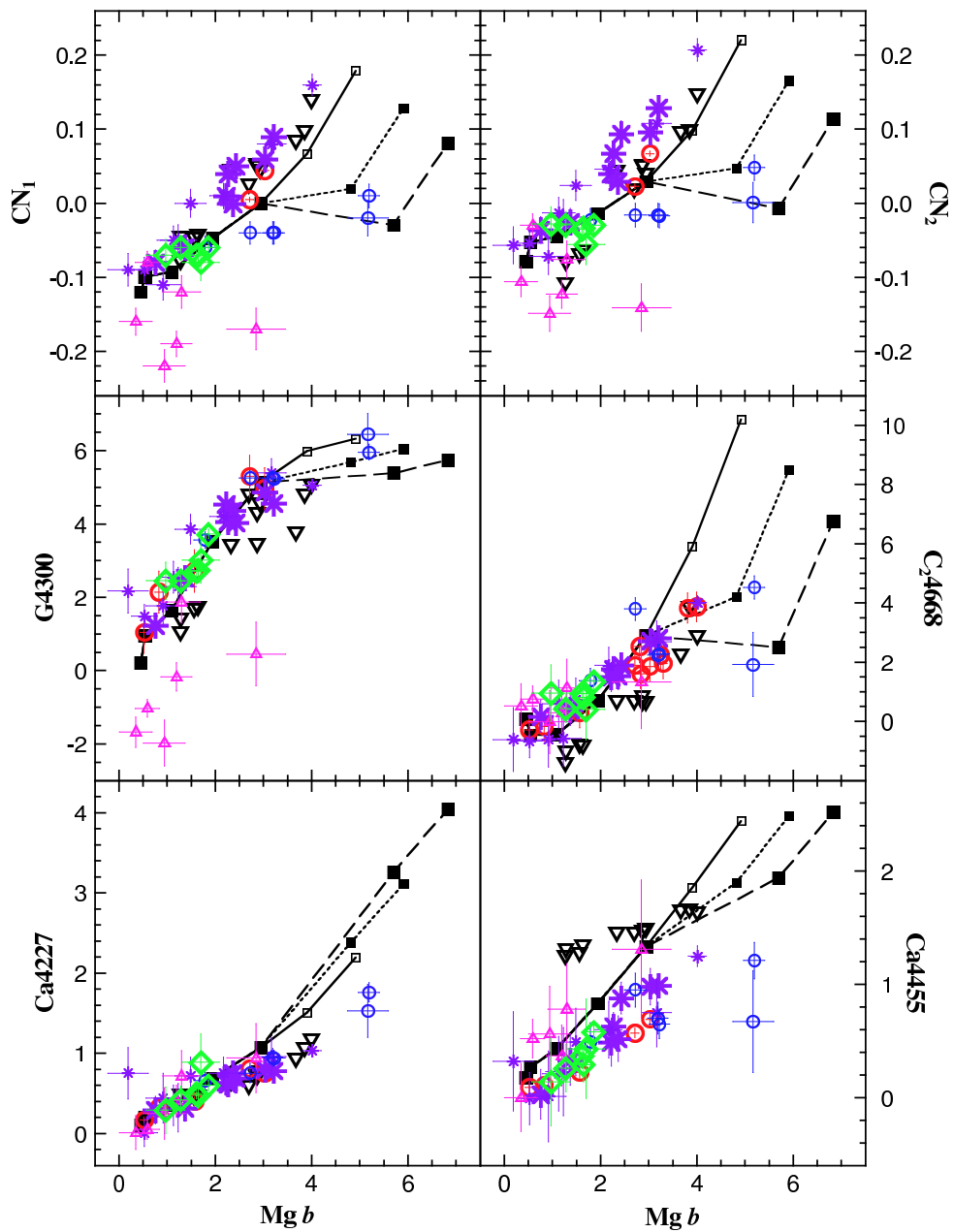}
\caption{Index-index diagrams Similar to Figure 3. 
M31 star clusters are compared with Milky Way GCs 
and 12 Gyr models for Mg $\it{b}$ vs. CN$_{1}$, CN$_{2}$, 
G4300, C$_{2}$4668, Ca4227, and Ca4455. Symbols are the same as 
in Figures 3 and 7.}
\end{figure}

\begin{figure}
\epsscale{0.8}
\plotone{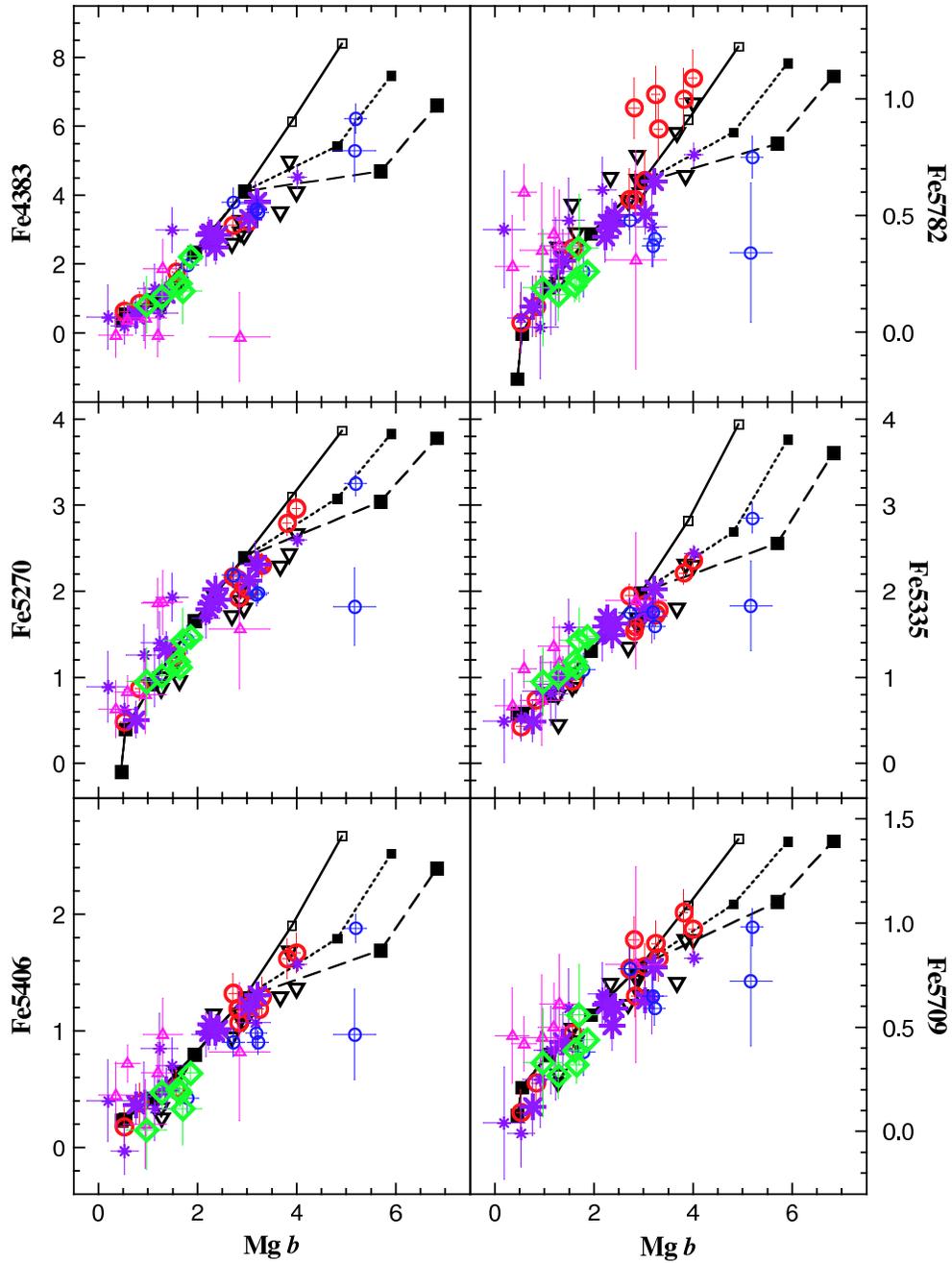}
\caption{Index-index diagrams similar to Figure 4. 
M31 star clusters are compared with Milky Way GCs 
and 12 Gyr models for Mg $\it{b}$ vs. Fe4383, Fe5782, Fe5270, 
Fe5335, Fe5406, and Fe5709. Symbols are the same as in Figures 3 and 7.}
\end{figure}

\begin{figure}
\epsscale{0.8}
\plotone{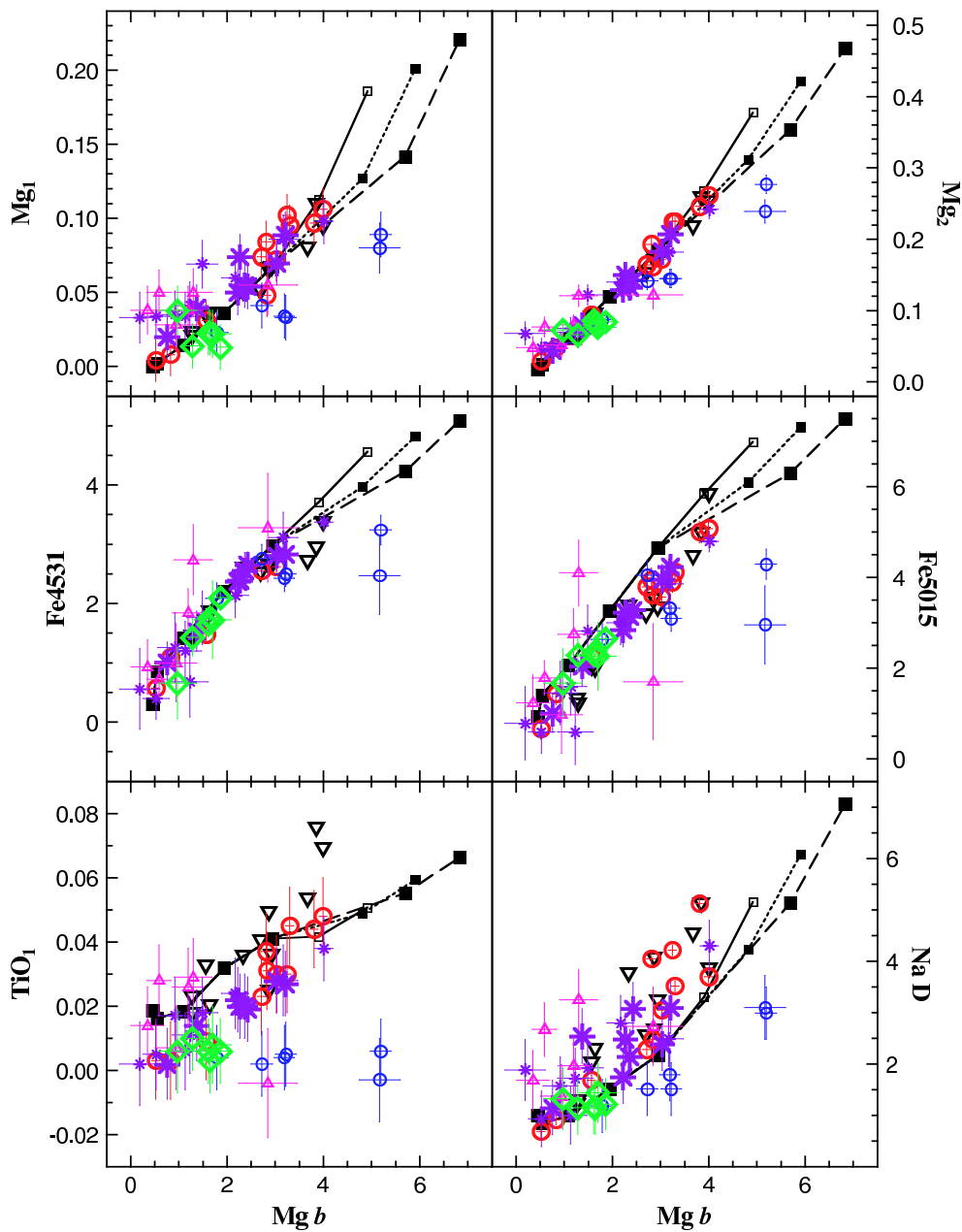}
\caption{Index-index diagrams similar to Figure 6. 
M31 star clusters are compared with Milky Way GCs 
and 12 Gyr models for Mg $\it{b}$ vs. Mg$_{1}$, Mg$_{2}$, 
Fe4531, Fe5015, TiO$_{1}$, and Na D. 
Symbols are the same as Figures 3 and 7.}
\end{figure}

\begin{figure}
\epsscale{0.7}
\plotone{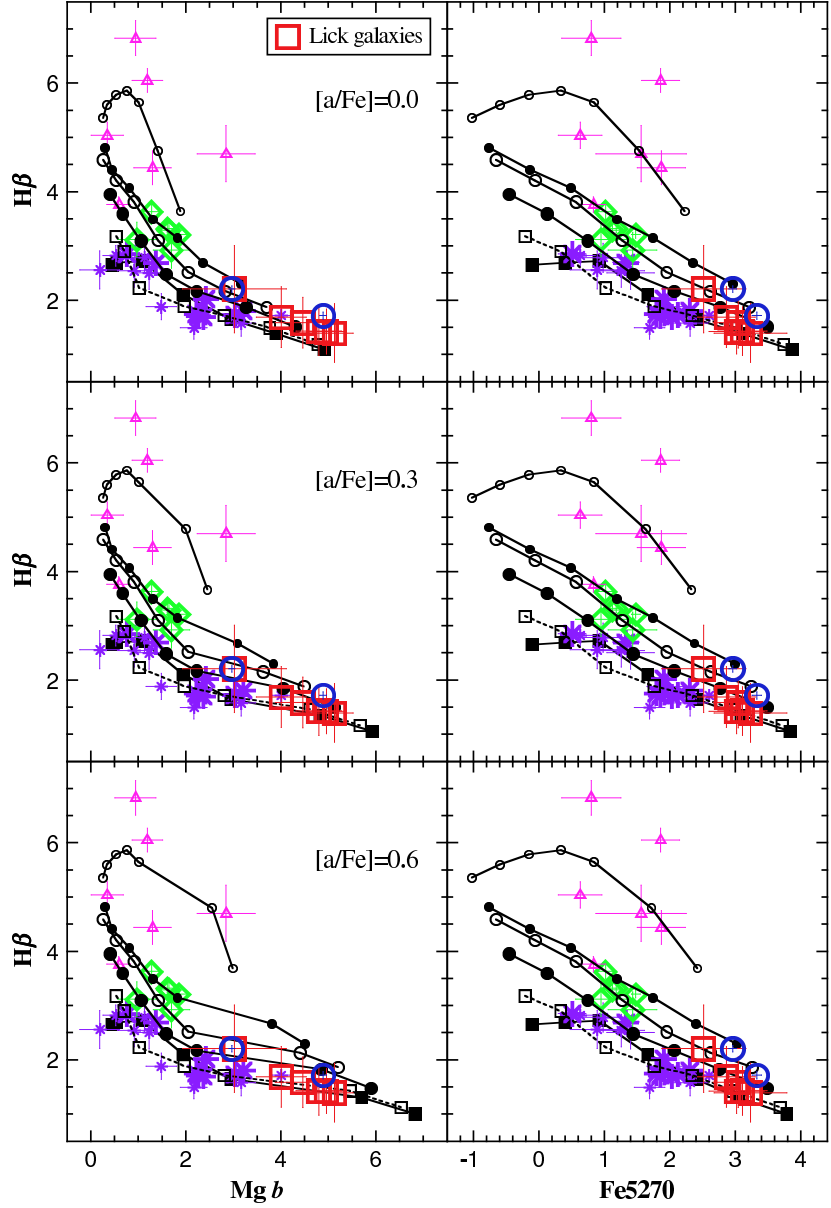}
\caption{Median Lick galaxies ({\rm big open squares}), M31 ({\em
right big open circle}), and M32 ({\em left big open circle}) are
compared with M31 star clusters (with symbols the same as Figure 7) and
with our models in Mg $\it{b}$ vs. H$\beta$ (left panel) and in Fe5270
vs. H$\beta$ (right panel). At given ages, HTWB02 response functions
are applied at [Fe/H] = 0.0 and +0.5 as indicated in the left panels.}
\end{figure}

\clearpage

\begin{figure}
\epsscale{0.7}
\plotone{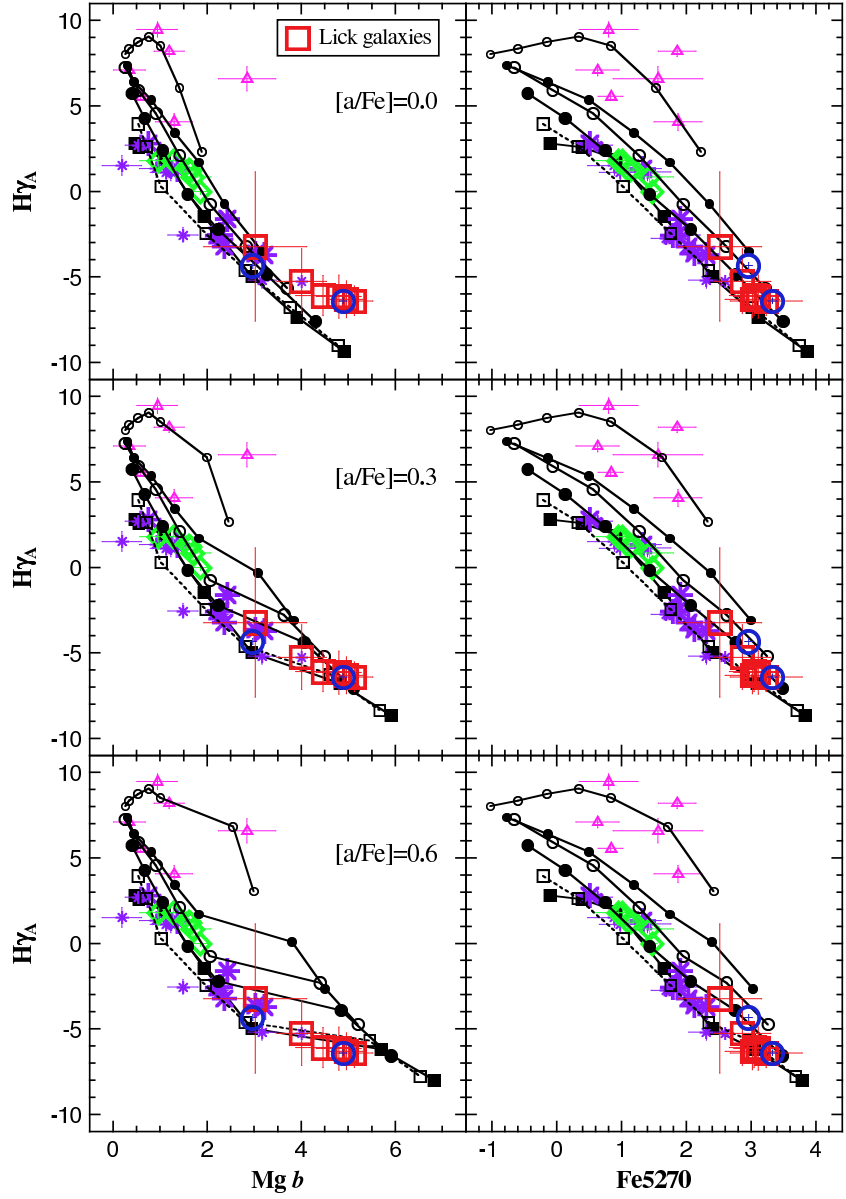}
\caption{Same as Figure 13, but in Mg $\it{b}$ vs. 
H$\gamma_{A}$ (left panel) and 
in Fe5270 vs. H$\gamma_{A}$ (right panel). Note that H$\gamma_{A}$ 
is significantly affected by $\alpha$-enhancement and results in 
widely different age and metallicity estimation, 
especially in the left panel.}
\end{figure}

\begin{figure}
\epsscale{0.8}
\plotone{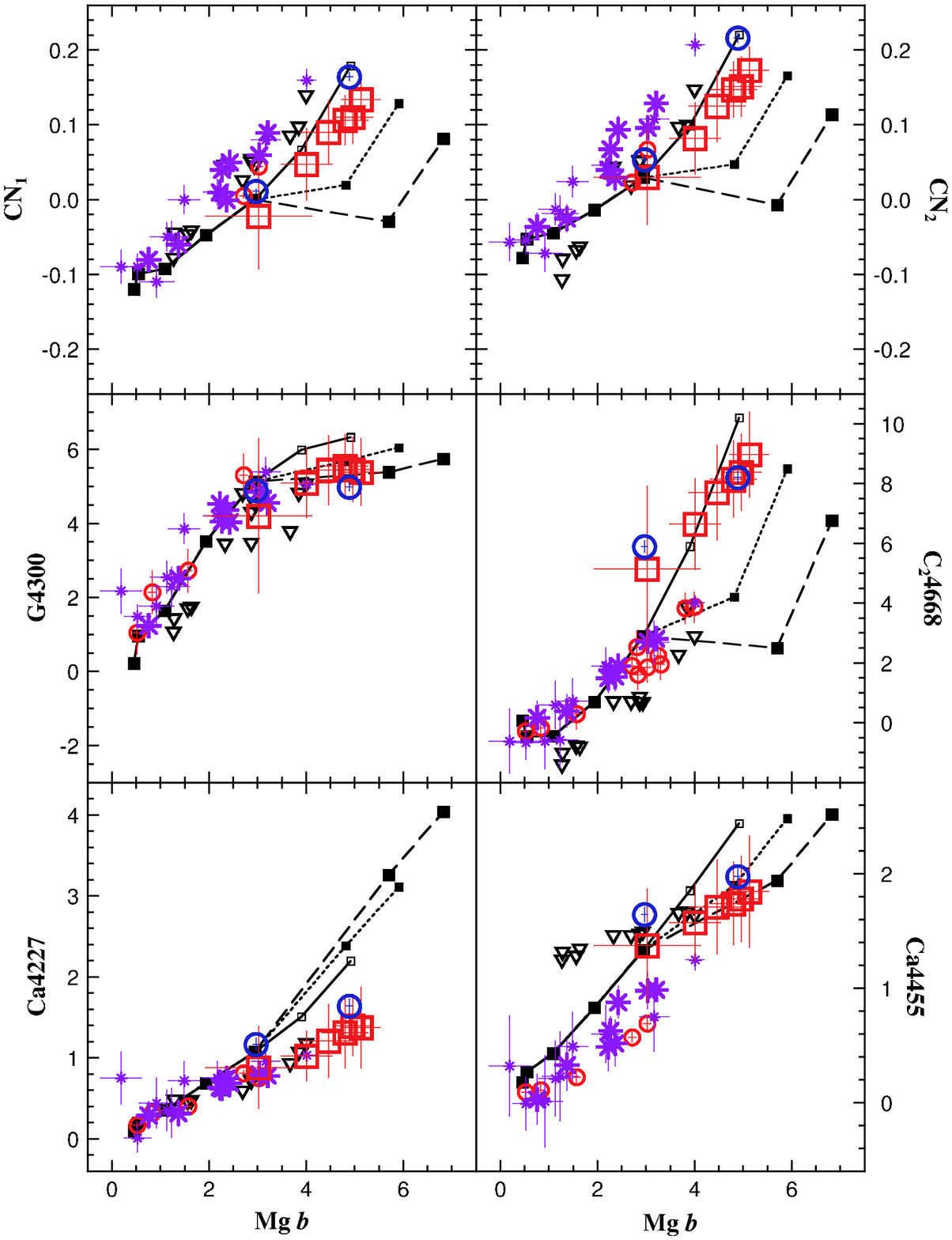}
\caption{Similar to Figure 10, but with
median Lick galaxies ({\rm big open squares}), M31 ({\em
right big open circle}), and M32 ({\em left big open circle})
compared with Milky Way GCs,  bona fide old M31 GCs, and 
12 Gyr models in Mg $\it{b}$ vs. CN$_{1}$, CN$_{2}$, 
G4300, C$_{2}$4668, Ca4227, and Ca4455. Cluster symbols are the same as 
in Figures 3 and 7.}
\end{figure}

\begin{figure}
\epsscale{1.0}
\plotone{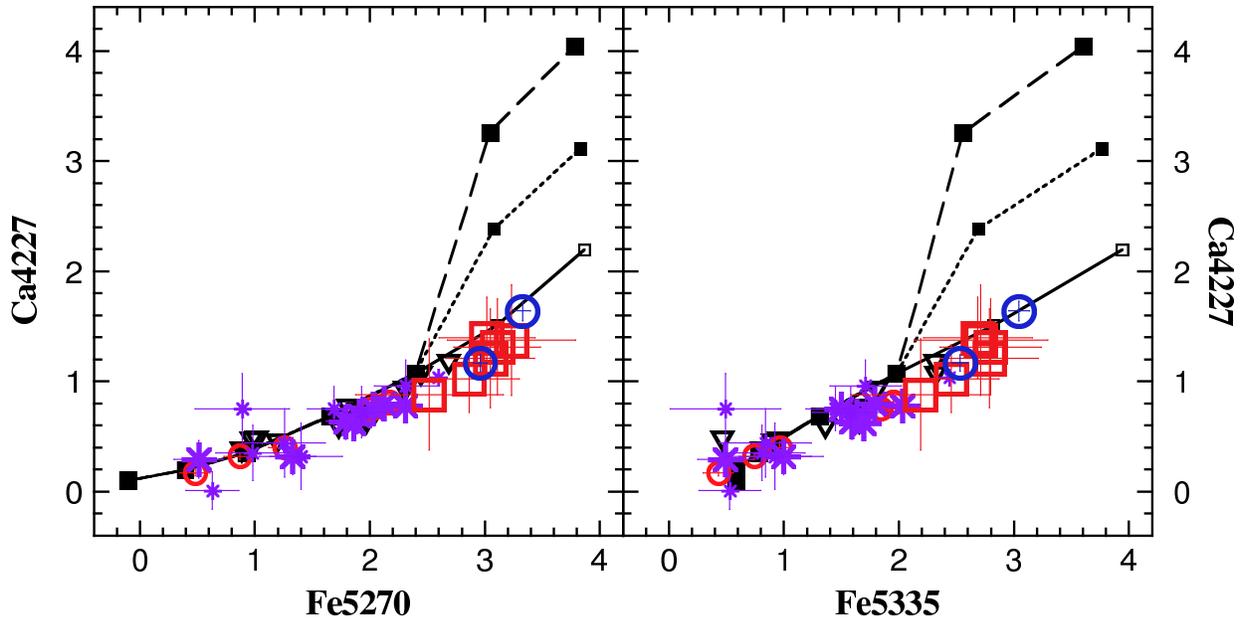}
\caption{Similar to the bottom left of Figure 15, but Ca4227 vs. 
Fe5270 (left) and Fe5335 (right). Note that the Ca underabundance is 
not outstanding in these plots.}
\end{figure}

\begin{figure}
\epsscale{0.8}
\plotone{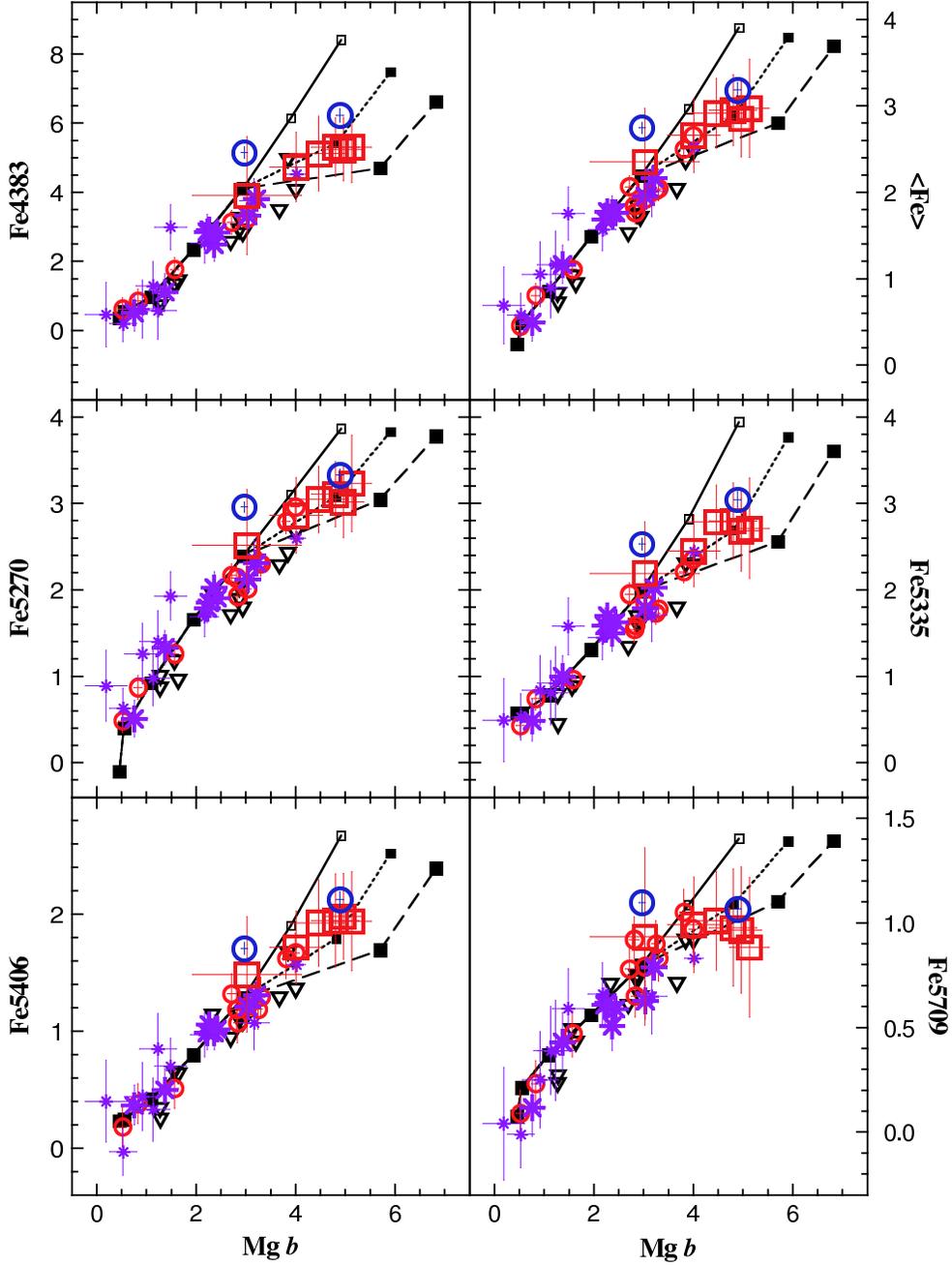}
\caption{Similar to Figure 11, but with
median Lick galaxies ({\rm big open squares}), M31 ({\em
right big open circle}), and M32 ({\em left big open circle})
compared with Milky Way GCs, bona fide old M31 GCs, and 
12 Gyr models 
in Mg $\it{b}$ vs. Fe4383, $<$Fe$>$, Fe5270, Fe5335, 
Fe5406, and Fe5709. Cluster symbols are the same as in Figures 3 and 7.}
\end{figure}

\begin{figure}
\epsscale{1.0}
\plotone{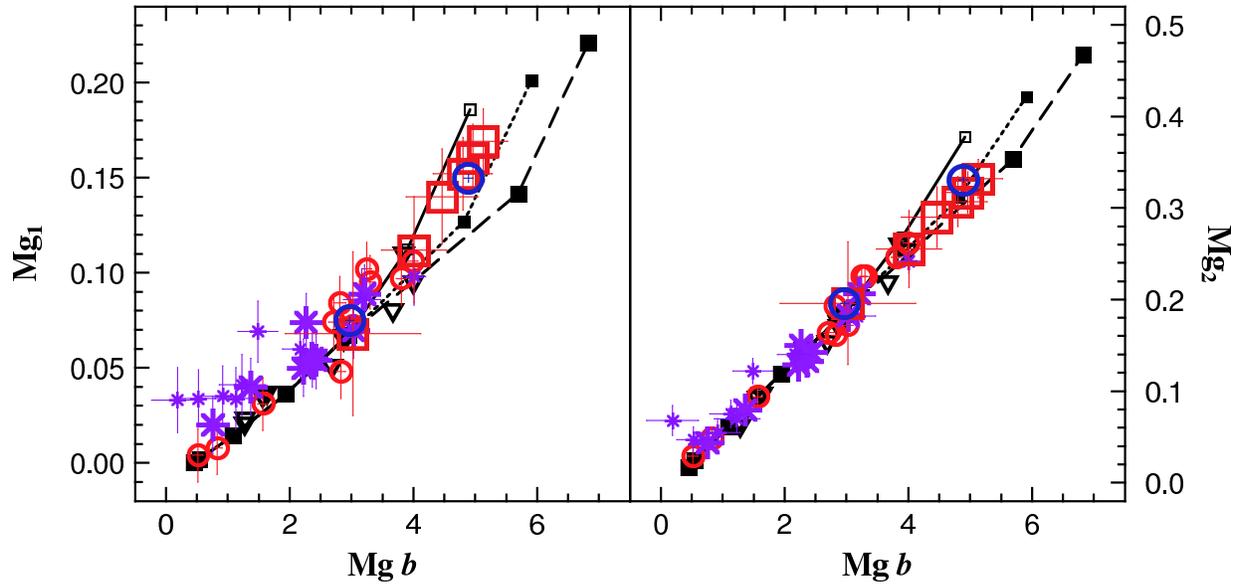}
\caption{Similar to Figure 5. 
Galaxies, Milky Way GCs, and bona fide old M31 GCs are compared
with 12 Gyr models in Mg $\it{b}$ vs. Mg$_{1}$ and Mg$_{2}$. 
Symbols are the same as in Figures 3, 7, and 13.}
\end{figure}

\begin{figure}
\epsscale{0.8}
\plotone{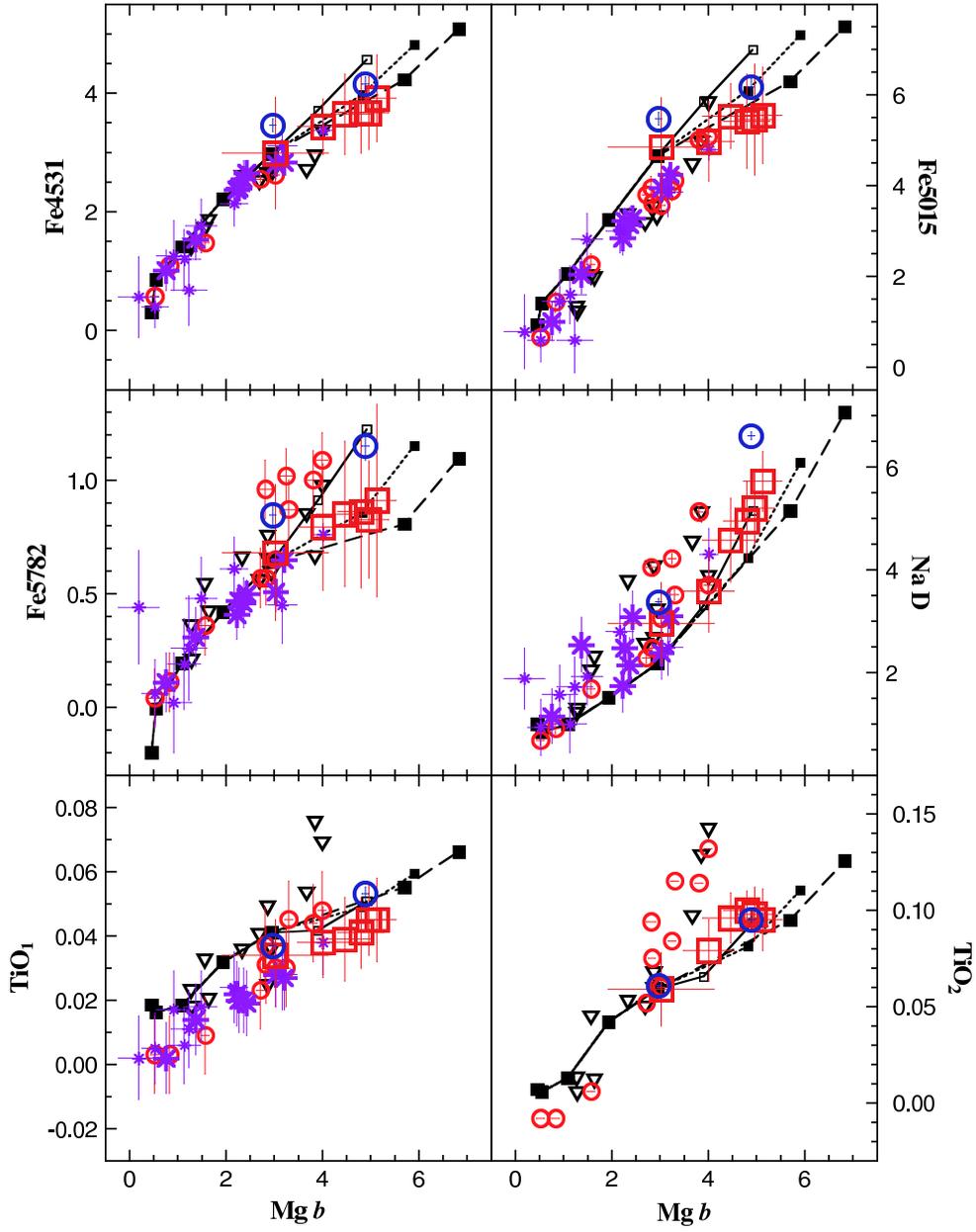}
\caption{Similar to Figure 12.
Galaxies, Milky Way GCs, and bona fide old M31 GCs are compared
with 12 Gyr models for Mg $\it{b}$ vs. Fe4531, Fe5015, Fe5782, 
Na D, TiO$_{1}$, and TiO$_{2}$. 
Symbols are the same as in Figures 3, 7, and 13.}
\end{figure}

\end{document}